\documentclass[useAMS,usenatbib]{mn2e}
\usepackage{hyperref}
\usepackage{times}
\usepackage{natbib}
\usepackage{aas_macros}
\usepackage{amsmath}
\usepackage{amssymb}
\usepackage[figuresright]{rotating}
\usepackage[labelfont=bf]{caption}
\usepackage{supertabular}
\usepackage{enumitem}

\newcommand{\ergs}{$\mathrm{erg\;s^{-1}}$}

\newcommand{\sar}{\lambda_\mathrm{sBHAR}}
\newcommand{\Sar}{$\sar$}
\newcommand{\msun}{\mathcal{M}_\odot}

\newcommand{\mstel}{\mathcal{M}_*}
\newcommand{\Mstel}{$\mstel$}
\newcommand{\lx}{L_\mathrm{X}}
\newcommand{\LX}{$\lx$}
\newcommand{\giv}{\;|\;}
\newcommand{\Psar}{$p(\log \sar \giv \mstel,z)$}

\newcommand{\PaperI}{\citetalias{Aird17}}

\newcommand{\upd}[1]{{#1}}
\newcommand{\updd}[1]{{#1}}
\newcommand{\refone}[1]{#1}

\begin{document}

\title[The distribution of AGN accretion rates]{X-rays across the galaxy population - II. The distribution of AGN accretion rates as a function of stellar mass and redshift}
\author[J. Aird et al.]{J. Aird$^{1}$\thanks{jaird@ast.cam.ac.uk}, A. L. Coil$^2$ and A. Georgakakis$^{3,4}$\\
$^1$Institute of Astronomy,
University of Cambridge,
Madingley Road,
Cambridge,
CB3 0HA\\
$^2$Center for Astrophysics and Space Sciences (CASS), Department of Physics, University of California, San Diego, CA 92093, USA\\
$^3$Max Planck Institute f\"{u}r Extraterrestrische Physik, Giessenbachstrasse, 85748 Garching, Germany\\
$^4$IAASARS, National Observatory of Athens, GR-15236 Penteli, Greece}
\pagerange{\pageref{firstpage}--\pageref{lastpage}} \pubyear{2017}

\maketitle
\label{firstpage}
\begin{abstract}
We use deep \textit{Chandra} X-ray imaging to measure the distribution of specific black hole accretion rates ($L_\mathrm{X}$ relative to the stellar mass of the galaxy) and thus trace AGN activity within star-forming and quiescent galaxies, as a function of stellar mass (from $10^{8.5}$--$10^{11.5}\mathcal{M}_\odot$) and redshift (to $z\sim4$).
We adopt near-infrared selected samples of galaxies from the CANDELS and UltraVISTA surveys, extract X-ray data for every galaxy, and use a flexible Bayesian method 
to combine these data and 
to measure the probability distribution function of specific black hole accretion rates, $\lambda_\mathrm{sBHAR}$.
We identify a broad  distribution of $\lambda_\mathrm{sBHAR}$  in both star-forming and quiescent galaxies---likely reflecting the stochastic nature of AGN fuelling---with a roughly power-law shape that rises toward lower $\lambda_\mathrm{sBHAR}$, a steep cutoff at 
$\lambda_\mathrm{sBHAR} \gtrsim0.1-1$ (in Eddington equivalent units), and a turnover or flattening at $\lambda_\mathrm{sBHAR} \lesssim10^{-3}-10^{-2}$. 
We find that the probability of a star-forming galaxy hosting a moderate $\lambda_\mathrm{sBHAR}$ AGN depends on stellar mass and evolves with redshift, shifting toward higher 
$\lambda_\mathrm{sBHAR}$ at higher redshifts.
This evolution is truncated at a point corresponding to the Eddington limit, indicating black holes may self-regulate their growth at high redshifts when copious gas is available.
The probability of a quiescent galaxy hosting an AGN is generally lower than that of a star-forming galaxy, shows signs of suppression at the highest stellar masses, and evolves strongly with redshift. 
The AGN duty cycle in high-redshift ($z\gtrsim2$) quiescent galaxies thus reaches $\sim$20 per cent, comparable to the duty cycle in star-forming galaxies of equivalent stellar mass and redshift. 
\end{abstract}

\section{Introduction}
\label{sec:intro}

The properties of galaxies and the supermassive black holes that lie at their centres appear to be fundamentally connected, but the physical origin of this relationship remains unclear \citep[see e.g.][]{Kormendy13}.
Supermassive black holes grow primarily through intense periods of mass accretion, during which they are observed as Active Galactic Nuclei (AGNs).
The total star formation rate density and the total AGN accretion density appear to follow similar patterns with cosmic time \citep[e.g.][]{Boyle98,Delvecchio14,Aird15}, indicating that the overall build up of galaxies and their central black holes proceeds in a coherent manner.
However, prior studies have found little evidence for a direct correlation between the current levels of black hole growth (traced by the AGN luminosity) and the wider-scale properties of a galaxy (such as the total stellar mass or the current rate of star formation, see e.g. \citealt{Azadi15} and references therein).
To determine the physical mechanisms that promote AGN activity---and determine the impact of AGN feedback on the evolution of galaxies---requires a detailed mapping of the distribution of AGN activity \emph{within} the evolving galaxy population.

X-ray surveys provide an efficient method of identifying AGNs over a wide range of redshifts and down to relatively low luminosities where the host galaxy dominates the observed light at other wavelengths \citep[see ][for a recent review]{Brandt15}.
A number of studies have thus taken X-ray selected samples of AGNs and examined the properties of the galaxies they lie in. 
Initial studies found that X-ray selected AGN samples predominantly lie in galaxies with red optical colours, indicating a possible connection between the presence of an AGN and the quenching of star formation throughout a galaxy \citep[e.g.][]{Nandra07b}. 
Subsequent studies, however, demonstrated that the majority of X-ray AGNs are found in galaxies with moderate-to-high stellar masses ($\mstel\gtrsim 3\times 10^{10}\msun$), 
although lower-mass hosts are also identified \citep[e.g.][]{Brusa09,Xue10}.
When compared to samples matched in stellar mass, X-ray AGNs either show no preference in terms of host colour or a mild preference to be found in star-forming galaxies \citep[e.g.][]{Silverman09b,Rosario13}.

\citet[hereafter A12]{Aird12}\defcitealias{Aird12}{A12} took a different approach, starting with a sample of galaxies out to $z\sim1$ 
\citep[from the Prism Multi-object Survey:][]{Coil11,Cool13} and determining the probability of finding an X-ray AGN \emph{within} such galaxies as a function of the galaxy properties: stellar mass, redshift, and colour. 
\citetalias{Aird12}~showed that the probability of hosting an AGN can be described by a power-law distribution of \emph{specific black hole accretion rates} (hereafter, \Sar, the rate of black hole accretion normalised relative to the host stellar mass).
The power-law distribution was found to be consistent over a broad range of stellar masses ($9.5\lesssim \log \mstel/\msun \lesssim 12$) but with a normalization that evolves strongly with redshift, indicating that the probability of hosting an AGN with a given \Sar\ is \emph{independent} of stellar mass but drops rapidly between $z\sim1$ and the present day. 
Thus, the predominance of higher mass host galaxies for samples of X-ray AGN can be ascribed to a selection effect: it is simply harder to identify AGNs with an equivalent \Sar\ in lower mass galaxies than in higher mass galaxies \citep[see also][]{Aird13}.  
\citet{Bongiorno12} confirmed these findings and extended the analysis to higher redshifts ($z\sim2.5$). 
\citetalias{Aird12} also showed that the probability of hosting an AGN (at a given \Sar) is mildly enhanced (by a factor $\sim2-3$) in galaxies with bluer optical colours.
This relationship was explored further by \citet{Azadi15}, who demonstrated an increase in the normalization of the distribution of \Sar\ with increasing star formation rate (measured relative to the ``main sequence of star formation": \citealt{Noeske07,Elbaz11}). 
\citet{Georgakakis14} also measured the distribution of \Sar\ for stellar-mass-limited samples of X-ray AGNs, split according to their rest-frame colours, confirming a broad distribution of \Sar\ and the dominance of AGNs in star-forming galaxies. 

These studies have had an important impact on our understanding of the relationship between galaxies and the growth of their central black holes. 
The broad distribution of accretion rates for samples of galaxies with similar physical properties indicates that the instantaneous level of AGN activity can vary over short timescales relative to the wider-scale properties of the galaxy \citep{Mullaney12b,Aird13,Hickox14,Schawinski15}.
Thus, the processes that bring gas into the centres of galaxies and trigger AGN activity appear to be stochastic in nature \citep[e.g.][]{Hopkins06c,Novak11}.
This variability of AGN activity can blur out any correlation between the overall levels of accretion and the galaxy properties, such as the star formation rate \citep[e.g.][]{Chen13,Stanley15}.
Thus, to determine the underlying relationship between AGN activity and galaxy evolution requires large, statistical samples of galaxies, enabling accurate measurements of the \emph{distribution} of accretion rates within different types of galaxies throughout cosmic history.

Despite this important progress, a number of open questions remain. 
A universal, power-law distribution of \Sar\ is difficult to reconcile with the overall shape and evolution of the total luminosity function of AGNs. 
A turnover at high \Sar\ (at a point roughly corresponding to the Eddington limit) and a steep power-law tail is required to recover the break in the X-ray luminosity function (at luminosity $L_*$) and the steep bright-end slope \citep[e.g.][]{Aird13,Caplar15}. 
Additional evolution in the \Sar\ distribution (e.g. a shift of any break to higher \Sar\ with increasing redshift) may also be required to recover the overall evolution of $L_*$ of the X-ray luminosity function \citep[e.g.][]{Aird10,Aird15}.  
\refone{Furthermore, some level of stellar-mass-dependent evolution of the \Sar\ distribution may be required to 
account for changes in the overall shape of the luminosity function toward higher redshifts \citep[e.g.][]{Aird13,Bongiorno16}.}
Ultimately, accurate measurements of the probability distribution of \Sar\ in different galaxy populations \emph{as a function of stellar mass and redshift} are required to build a coherent picture of galaxy--AGN co-evolution.

In this paper, we build on the approach pioneered by \citetalias{Aird12} to measure the distribution of accretion rates within the galaxy population.
We construct a large, stellar-mass-limited sample of $\sim120,000$ galaxies spanning out to $z\sim4$ and with a wide range of stellar masses ($8.5\lesssim \log \mstel/\msun\lesssim 11.5$).
We extract X-ray data from deep \textit{Chandra} imaging for every galaxy in our sample and use a sophisticated Bayesian methodology to probe below the nominal detection limits of the \textit{Chandra} imaging and recover the intrinsic underlying distribution of \Sar.
Our method is non-parametric and allows us to divide our galaxy sample according to redshift, stellar mass, and galaxy type (star-forming and quiescent) and track how the incidence of AGNs and the distribution of their accretion rates changes across the evolving galaxy population.
Our approach, starting with a large \emph{galaxy} sample, contrasts with other recent studies that start with an X-ray selected sample of AGNs, determine the properties of their host galaxies, and use either parametric \citep{Bongiorno16} or non-parametric \citep{Georgakakis17} methods to reconcile the observed X-ray AGN population with independent measurements of the galaxy stellar mass function and thus recover the underlying distribution of \Sar\ \citep[see also][]{Aird13}. 

This paper is the second in a series, building on the analysis of \citet[hereafter Paper I]{Aird17}.\defcitealias{Aird17}{Paper~I} 
In \PaperI, we measured the intrinsic distribution of X-ray luminosities for samples of star-forming galaxies, binned by stellar mass and redshift.
In each bin we identified two different origins for the X-ray emission:  
1) the total emission from X-ray binaries and hot gas, tracing the average star formation rate of our galaxy sub-samples; 
and
2) the emission from active galactic nuclei (AGN), tracing supermassive black hole accretion activity.
In this paper (hereafter Paper II) we focus on the AGN emission. 
We present measurements of the intrinsic distributions of \Sar\
within all galaxies, as well as within the star-forming and quiescent galaxy populations, as a function of stellar mass and redshift.
We compare the shapes of these distributions and analyse their evolution with redshift and their stellar mass dependence, providing crucial insights into the extent and distribution of black hole growth across the galaxy population. 
\updd{
Paper III (Aird et al. in preparation) will expand this analysis further, presenting measurements of the distribution of \Sar\ within galaxies as a function of star formation rate (relative to the main sequence of star formation) and thus exploring the connection between the growth of galaxies and their central black holes in greater detail.}

Section~\ref{sec:data} briefly describes our data sets and the sample selection for this paper. 
In Section~\ref{sec:accdists} we present our measurements of the probability distribution functions of \Sar\ as a function of stellar mass and redshift for all galaxies (Section~\ref{sec:all}) as well as within the star-forming and quiescent populations separately (Section~\ref{sec:sfqu}). 
We also use these distributions to estimate the AGN duty cycle (the fraction of galaxies with black holes growing above a certain \Sar) and the average accretion rate of these active black holes, providing further insights into the dependence of AGN activity on stellar mass and redshift (Section~\ref{sec:dutycycle}).
Section~\ref{sec:discuss} discusses our findings and we summarize our overall conclusions in Section~\ref{sec:conclusions}. 
We adopt a flat cosmology with $\Omega_\Lambda = 0.7$ and $H_0 = 70$~km~s$^{-1}$~Mpc$^{-1}$ throughout this paper and assume a \cite{Chabrier03} stellar initial mass function (IMF).

\section{Data and sample selection}
\label{sec:data}

\begin{figure*}
\includegraphics[width=\textwidth,trim=20 30 0 40]{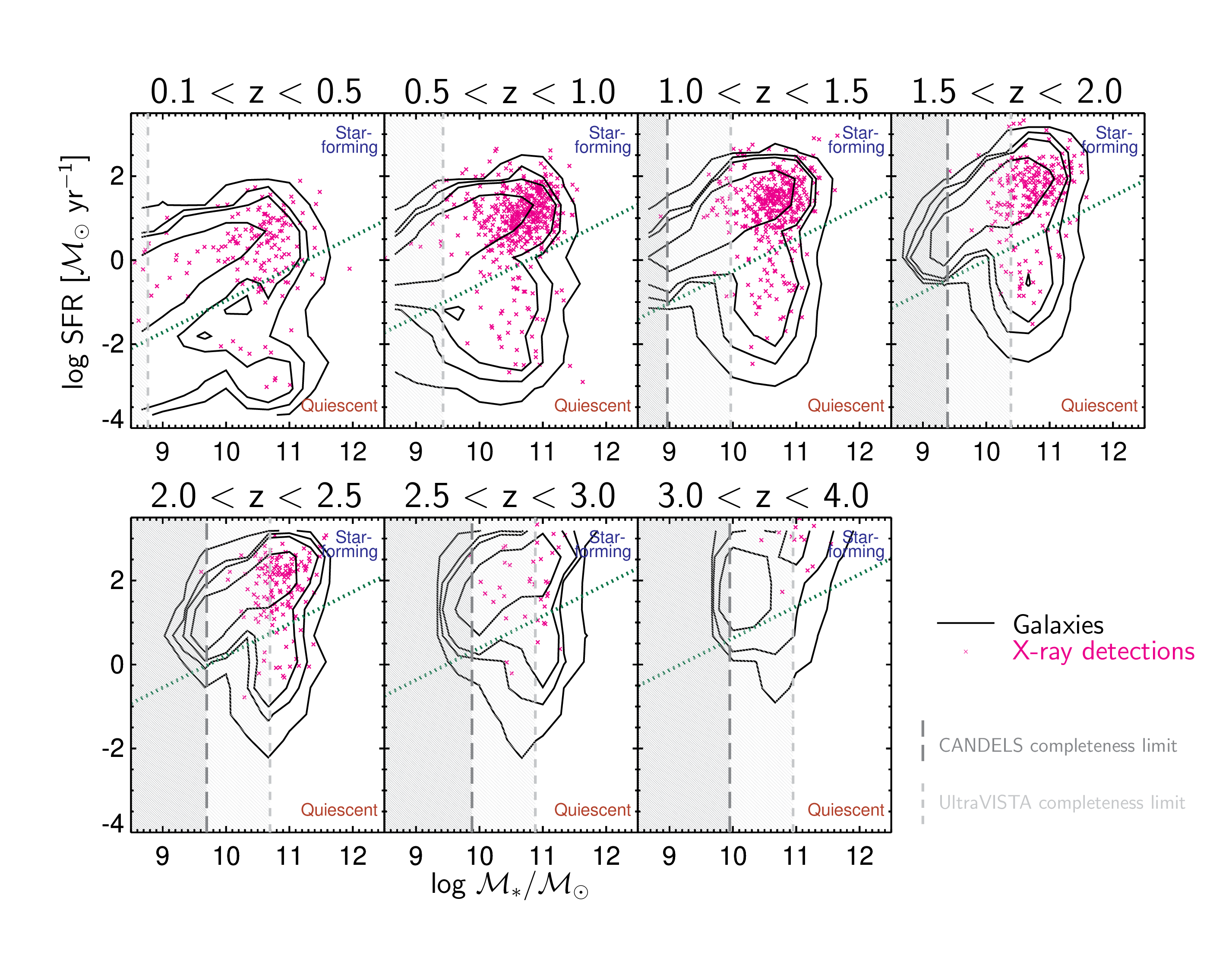}
\caption{
UV-to-IR based SFR estimates (either summing the UV+IR light or based on UV-to-NIR SED fits) versus stellar mass (from SED fitting) at different redshifts for our near-infrared selected galaxy sample from CANDELS/3DHST and UltraVISTA (black contours, corresponding to 68, 90, 95 and 99 per cent of the sample). 
Pink crosses indicate hard X-ray selected AGNs (SFRs are corrected for AGN contamination, although AGN-dominated sources are excluded from this plot; see Appendix~\ref{app:agnfast}).
\upd{The grey shaded regions indicate our stellar mass completeness limits (evaluated at the centre of the redshift bin), with the dark-grey long-dashed line showing the limit for the deep CANDELS imaging and the light-grey dashed line showing the limit for the wider-area UltraVISTA imaging.}
We divide our galaxies into ``star-forming" (above green dotted line) and ``quiescent" (below green dotted line) galaxies on the basis of their SFRs relative to the evolving main sequence (taken from Paper~I, see Equation~\ref{eq:sfms_cut}).
}
\label{fig:sfr_vs_mstel}
\end{figure*}

The results in this paper are based on the same datasets used in Paper~I, described fully therein and briefly summarized here.

Our work is based on near-infrared (NIR) selected samples of galaxies taken from four of the CANDELS survey fields \citep[GOODS-S, GOODS-N, AEGIS and COSMOS:][]{Grogin11,Koekemoer11} and the larger UltraVISTA survey of $\sim$1.6~deg$^2$ of the COSMOS field \citep{McCracken12}. We adopt photometric catalogues from \citet{Skelton14} and \citet{Muzzin13} for CANDELS and UltraVISTA respectively.

We match the NIR-selected catalogues with X-ray detected sources in the \textit{Chandra} imaging of our fields \citep{Alexander03,Xue11,Nandra15,Civano16}, where all the \textit{Chandra} data have been reprocessed using our own procedures \citep[see][]{Laird09,Georgakakis14,Nandra15,Aird15}. 
We also extract X-ray data (total counts, background counts, and effective exposures) at the positions of the remaining NIR-selected sources that are not matched to a significant X-ray detection,
\refone{We extract counts within a circular aperture that corresponds to a 70 per cent enclosed energy fraction for the exposure-weighted \textit{Chandra} point spread function at each source position.}
These extracted X-ray data contain information on the sensitivity of the \textit{Chandra} imaging as well as providing information from sources that fall below the nominal X-ray detection threshold that is used in our analysis (see Section~\ref{sec:accdists} and Appendix~\ref{app:updatedbayes} below). 

In this paper, we use X-ray information from the hard (2--7~keV) band rather than the 0.5--2~keV band adopted in Paper~I.
While \textit{Chandra} is less sensitive in the 2--7~keV band, using a harder band reduces biases due to intrinsic absorption and allows us to reliably estimate the X-ray luminosity of AGNs, assuming absorption columns of $N_\mathrm{H}\lesssim10^{23}$~cm$^{-2}$.
\refone{We estimate intrinsic rest-frame 2--10~keV X-ray luminosities based on the observed count rates in the 2--7~keV band.
Following \citet{Aird15}, to convert the observed count rates to an X-ray luminosity we assume an X-ray spectrum with a fixed photon index of $\Gamma=1.9$, a reflection component (using the \textit{pexrav} model) with fixed strength of $R=1$, and Galactic absorption only.} 
We note that the impact of \emph{intrinsic} absorption on the observed 2--7~keV band depends on redshift and is thus reduced at $z\gtrsim1$, although we will still severely underestimate the luminosities of Compton-thick sources ($N_\mathrm{H}\gtrsim10^{24}$~cm$^{-2}$). 
A full and accurate characterisation of intrinsic absorption---including the resulting selection biases---requires advanced techniques \citep[e.g.][]{Buchner15,Aird15} and is beyond the scope of this paper.

We compile high-resolution spectroscopic redshifts \citep[see][and references therein for full details]{Skelton14,Muzzin13,Aird15,Aird17,Marchesi16}, low-resolution spectroscopic redshifts (i.e. from PRIMUS, \citealt{Coil11}, or 3DHST, \citealt{Momcheva16}) and photometric redshifts \citep[including high quality AGN photo-$z$, where available, from][]{Hsu14,Nandra15,Marchesi16} for sources in our NIR-selected catalogues.
Rest-frame colours are calculated using \textit{EaZY} \citep{Brammer11}.
We use the FAST code \citep{Kriek09} to fit the UV-to-NIR spectral energy distributions (SEDs) of our sources, as described in the appendix of \PaperI. 
Appendix~\ref{app:agnfast} of this paper (see below) describes our modifications of FAST to account for an AGN component in addition to a galaxy component when fitting the SEDs.
We only allow for an AGN component for objects with significant X-ray detections.
We also estimate star-formation rates (SFRs) based on either the sum of the UV and IR emission (for sources with 24\micron\ detections) or the dust-corrected SFR estimate from the SED fit (see \PaperI\ for details). 
These SFRs are corrected for any AGN contribution as described in Appendix~\ref{app:agnfast} below.

To define our sample of galaxies, we apply the magnitude limits and the stellar-mass completeness limits described in \PaperI. 
While \PaperI\ focused on star-forming galaxies, in this paper we measure the distribution of AGN activity across the entire galaxy population as a function of stellar mass, redshift, and galaxy type.
We classify galaxies as star-forming or quiescent based on their SFRs relative to the evolving ``star-forming main sequence". 
Figure~\ref{fig:sfr_vs_mstel} shows the distribution of SFR 
and \Mstel\ 
for our galaxy sample at different redshifts (pink crosses correspond to directly detected X-ray sources). 
A star-forming main sequence with a roughly constant slope and a normalization that increases toward higher redshifts is apparent in our samples out to at least $z\sim3$ and is reasonably well described by the relation given by equation 6 in Paper~I. 
We define a cut (shown by the green dotted lines in Figure~\ref{fig:sfr_vs_mstel}) at 1.3~dex below the evolving main sequence from Paper~I, which roughly corresponds to the minimum in the distribution of SFRs relative to the main sequence and is given by
\begin{equation}
\log \mathrm{SFR_{cut}} \mathrm{[\msun yr^{-1}]} = -8.9 + 0.76\log \dfrac{\mstel}{\msun} + 2.95 \log (1+z).
\label{eq:sfms_cut}
\end{equation}
We classify galaxies that fall below this cut as quiescent; those above the green dotted line are classified as star-forming galaxies.
\upd{The vertical grey dashed lines in Figure~\ref{fig:sfr_vs_mstel} indicate the stellar mass completeness limits we apply for the CANDELS and UltraVISTA samples. 
We note that the majority of quiescent galaxies in our sample are identified in the larger-area (but shallower) UltraVISTA imaging; the paucity of lower-mass quiescent galaxies at high-redshifts in our sample is due to a combination of the \emph{intrinsic} rarity of such galaxies and the lack of data to the required (i.e. CANDELS) depth over a sufficiently large area.} 

\upd{
Sources  identified as ``AGN-dominated" at optical to IR wavelengths based on our two-component SED fitting (i.e. where more than 50 per cent of the light at rest-frame 5000\AA\ is from the AGN) are included in our sample of all galaxies.
For the majority of these sources ($\sim77$~per cent) the galaxy template dominates the emission at $\sim1$\micron\ (in our best fit), indicating that we can still estimate the stellar mass of the host galaxy to a reasonable degree of accuracy.  
However, we exclude all AGN-dominated sources (which constitute $\lesssim 20$ per cent of the X-ray detected samples and $\lesssim0.3$~per cent of the overall galaxy sample) when considering the star-forming or quiescent samples separately as it is not possible to reliably measure the SFR and thus classify these sources into either group.
In Appendix~\ref{app:agndom} we show that excluding the AGN-dominated sources from this analysis has a negligible impact on our overall results and conclusions. 
}

Our final sample consists of 126,971 galaxies (above our stellar-mass completeness limits), of which 106,636 are classified as star-forming, 19,971 as quiescent, and 364 as AGN-dominated (precluding a classification of the host galaxy type). 
A total of 1,797 sources in our sample are directly detected in the 2--7~keV X-ray band \upd{(including 318 of the sources classified as AGN-dominated)}.

\section{The distribution of AGN accretion rates across the galaxy population}
\label{sec:accdists}

In this section, we present measurements of the distribution of AGN accretion rates as a function of galaxy type, stellar mass and redshift.
In Section~\ref{sec:method} we give a brief description of our methodology (further details are given in Appendix~\ref{app:updatedbayes}). 
Section~\ref{sec:all} presents measurements of the distributions of specific black hole accretion rates for all galaxies as a function of stellar mass and redshift, while Section~\ref{sec:sfqu} presents measurements for the star-forming and quiescent populations separately. 
In Section~\ref{sec:dutycycle} we calculate the AGN duty cycle and the average accretion rate (above a fixed limit), summarizing 
our results on how the accretion rate distributions depend on host galaxy properties.

\subsection{Methodology}
\label{sec:method}

Following \citetalias{Aird12}, throughout this paper we measure the distribution of \emph{specific black hole accretion rates} (\Sar), the rate of accretion \upd{onto the central supermassive black hole} scaled relative to the stellar mass of the host galaxy.
Defining AGN activity in this manner allows us to trace the wide distribution in accretion rates within galaxies of fixed stellar mass.
We also account for the bias whereby a higher mass galaxy with a given \Sar\ appears more luminous than a lower mass galaxy that is growing its central black hole at a similar rate. 
Based on the same approach, \citetalias{Aird12} found that at $z\lesssim1$ the incidence of AGN activity was defined by a single, power-law distribution of \Sar\ across a wide range of galaxy stellar masses ($\mstel \sim 10^{9.5-12}\msun$), indicating that \Sar\ may be a more fundamental property of AGNs than the directly observed luminosity.

We define \Sar\ in dimensionless units, such that
\begin{equation}
\sar  = \frac{ k_\mathrm{bol} \; L_\mathrm{X} }
                  { 1.3 \times 10^{38} \; \mathrm{erg\;s^{-1}} \times 0.002 \dfrac{\mathcal{M}_*}{\mathcal{M}_\odot} }
\label{eq:sar}
\end{equation}
where $L_\mathrm{X}$ is the rest-frame 2--10~keV X-ray luminosity, $k_\mathrm{bol}$ is a bolometric correction factor (we adopt a constant $k_\mathrm{bol}=25$ in this paper) and $\mathcal{M}_*$ is the total stellar mass of the AGN host galaxy (estimated from our SED fitting).
The additional scale factors are chosen such that $\sar \approx \lambda_\mathrm{Edd}$, the Eddington ratio, assuming that the mass of the central black hole scales directly with the total host stellar mass. Thus, the \Sar\ distribution can be regarded as a tracer of the distribution of Eddington ratios. 
However, these scaling relations are expected to have significant intrinsic scatter and may break down in certain parameter regimes (e.g. at lower stellar masses). 
Nonetheless, \Sar\ provides a meaningful measurement of the rate at which a given galaxy is growing its black hole, relative to the mass of the galaxy. 
With a fixed $k_\mathrm{bol}$, with our definition $\sar \propto L_\mathrm{X}/\mstel$ \citep[see][]{Bongiorno12,Bongiorno16}.

To trace the distribution of \Sar\, we measure the probability density function of $\log \sar$ at a given stellar mass and redshift, $p(\log \sar \giv \mstel,z)$.
\upd{
Our measurement of $p(\log \sar \giv \mstel,z)$ describes the probability that a galaxy with a given stellar mass and redshift hosts an AGN with a given specific accretion rate.
Equivalently, 
this quantity reflects the \emph{distribution} of specific accretion rates within a sample of galaxies in a fixed range of stellar mass and redshift.
}
We adopt the Bayesian mixture modelling approach described in appendix~B of Paper~I,
adapted to consider \Sar\ rather than $L_\mathrm{X}$ (see Appendix~\ref{app:updatedbayes} below for more details). 
Our method provides a flexible method of recovering the underlying distribution that does not assume a particular functional form and has relatively few constraints, other than a prior that prefers a smoothly varying distribution and a requirement that the total probability distribution function must integrate to less than 1. 

Our method accounts for uncertainties in individual X-ray flux measurements and the varying sensitivity of the \textit{Chandra} X-ray imaging over the CANDELS and UltraVISTA fields.
\upd{
The variation in the X-ray sensitivity of the deep \textit{Chandra} surveys---varying both between and \emph{within} our fields---is extremely well characterized and is driven by the vignetting of the telescope and degradation of the point spread function with off-axis angle, in addition to the overall exposure time.  
Our statistical approach uses the available X-ray information for \emph{every} galaxy in our sample, including the observed counts and the sensitivity information at the position of galaxies that fall below the nominal X-ray detection threshold.
By using the available X-ray information from every galaxy in our sample we are able to place much tighter constraints on \Psar. 
In appendix C of \PaperI\ we performed a number of simulations to verify that our method is able to combine the variable-depth \textit{Chandra} data and 
accurately recover the shape of different underlying distributions.}

At the lowest X-ray luminosities probed by our data, galactic non-AGN processes (high- and low-mass X-ray binaries, hot gas, supernovae) can make a significant contribution to the observed luminosity, as explored in \PaperI. 
We have thus updated our statistical method to allow for the expected contribution from galactic processes to the observed X-ray counts, based on our multiwavelength measurements of SFR and \Mstel\ for each galaxy in our sample.
We correct for this potential contamination when recovering the distribution of AGN accretion rates, $p(\log \sar \giv \mstel,z)$.
The updates to our statistical method are described in Appendix~\ref{app:updatedbayes} below.
However, our estimates of $p(\log \sar \giv \mstel,z)$ remain highly uncertain in regimes where the average galactic emission dominates over the X-ray AGN signal.
In Figure~\ref{fig:pledd_all} (described fully in Section~\ref{sec:all} below), we use dashed and dotted lines to indicate regimes where \Sar\ corresponds to a luminosity within 0.5~dex of the ``X-ray main sequence of star formation", identified in Paper~I.
Our measurements of $p(\log \sar \giv \mstel,z)$ should be treated with caution in these regimes.
In subsequent plots we only consider results above these limits, where we can reliably trace the distribution of AGN emission.

\subsection{The distribution of accretion rates within all galaxies as a function of stellar mass and redshift}
\label{sec:all}

\begin{figure*}
\includegraphics[width=\textwidth,trim=0 0 0 -20]{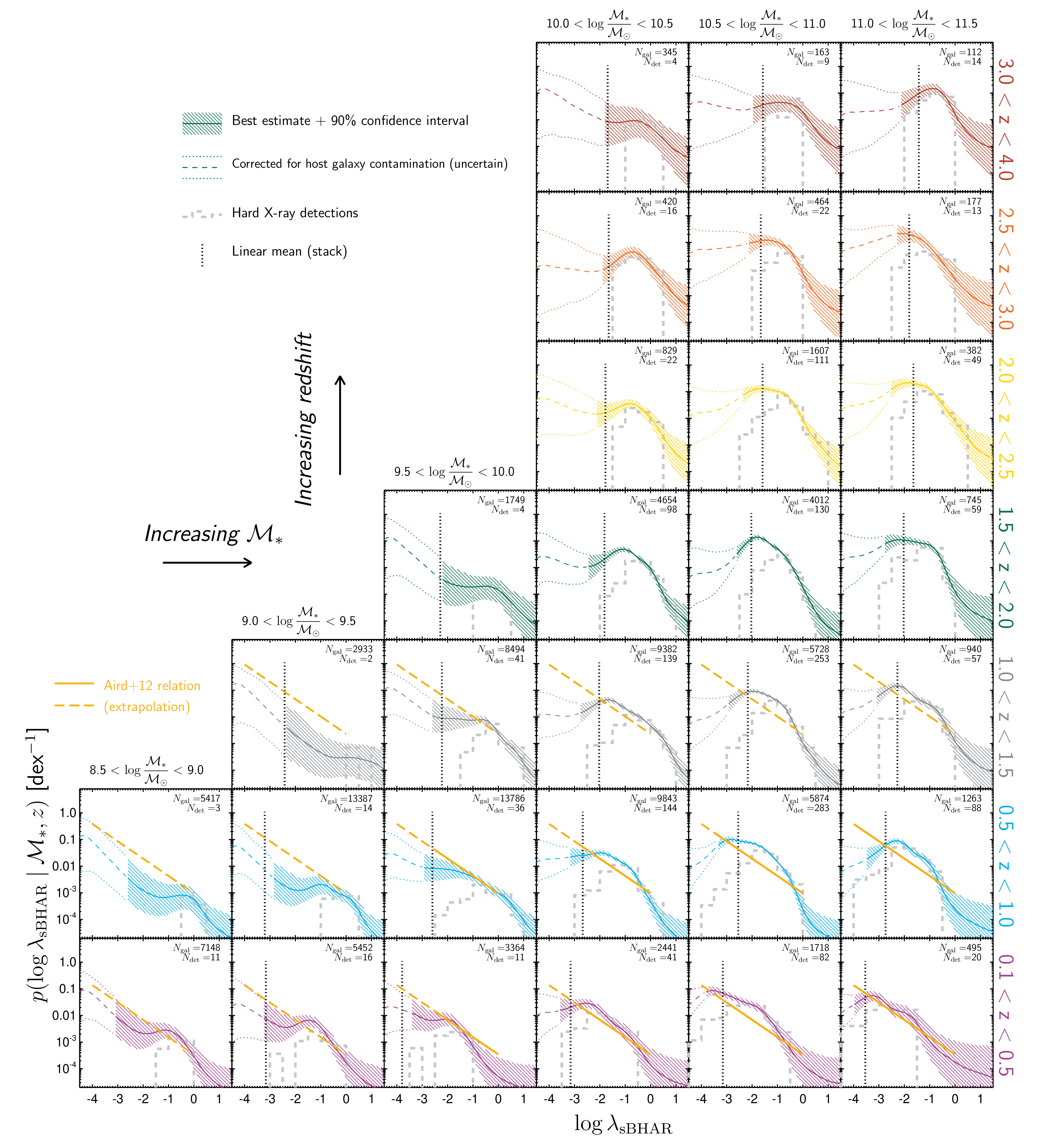}
\caption{
The probability distribution of specific black hole accretion rates, \Sar, for all galaxies (both star-forming and quiescent) as a function of stellar mass (increasing to the right) and redshift (increasing to the top). 
The thick coloured lines show the best estimate of $p(\log \sar \giv \mstel,z)$ using our flexible Bayesian mixture modelling approach (shaded regions give the 90 per cent confidence intervals). 
At low \Sar\ we have corrected for the contribution from galactic (non-AGN) X-ray emission, thus our estimates are less certain in these regimes, indicated by the dashed and dotted lines.
At low redshifts, the orange line indicates the power-law function for $p(\sar \giv \mstel,z)$ measured by \citet{Aird12} which is independent of stellar mass but evolves with redshift. The dashed orange lines indicate a mild extrapolation of this relation to regimes that were not directly probed in \citet{Aird12}. 
The grey dashed histograms show the observed distributions of \Sar\ for galaxies that satisfy the nominal X-ray detection criterion in the 2--7~keV band.
The total number of galaxies and the total number of hard X-ray detections in each bin is given in the legend of each panel.  
}
\label{fig:pledd_all}
\end{figure*}

Figure~\ref{fig:pledd_all} presents measurements of $p(\log \sar \giv \mstel,z)$ for our full galaxy sample as a function of stellar mass and redshift. 
The thick coloured lines and hatched regions show our best estimates of $p(\log \sar \giv \mstel,z)$ and the 90 per cent confidence intervals, respectively, as recovered by our Bayesian modelling. 
The grey dashed histograms indicate the observed distribution of \Sar\ based on the X-ray detected population only without any corrections for the X-ray incompleteness, \upd{which differ significantly from our estimates of the intrinsic distributions at lower values of \Sar\ that use the available X-ray data and sensitivity information from \emph{all} galaxies in a given sample (see Section \ref{sec:method} above). 
}

The results shown in Figure~\ref{fig:pledd_all} reveal the wide distribution of \Sar, rising towards lower \Sar\ and spanning at least $2-3$ orders of magnitude, within galaxies of all stellar masses and redshift ranges probed here. 
Generally, the distributions have a relatively steep cutoff at the highest accretion rates ($\log \sar \gtrsim -1$, but dependent on stellar mass and redshift), a flatter distribution at moderate accretion rates ($-3 \lesssim \log \sar \lesssim -1$), and in some cases a turnover towards lower \Sar\ resulting in a very broad peak.
The orange lines in the low-redshift ($z<1.5$) panels of Figure~\ref{fig:pledd_all} indicate the power-law relation measured by \citetalias{Aird12} (dashed lines indicate an extrapolation of the measured relation), which is independent of stellar mass but has a normalization that evolves with redshift. 
While this relation is broadly consistent with our measurements, our updated analysis reveals that the accretion rate distribution has a more complicated shape.
There is also clear evidence of a stellar-mass dependence, at least at $z>0.5$, whereby the probability of hosting an AGN of $-3 \lesssim \log \sar \lesssim -1$ is higher than the \citetalias{Aird12} relation at high stellar masses ($\log \mstel/\msun \gtrsim 10.5$) and is below the extrapolation at low stellar masses ($\log \mstel/\msun \lesssim 9.5$).

\begin{figure*}
\includegraphics[width=\textwidth,trim=0 0 0 -30]{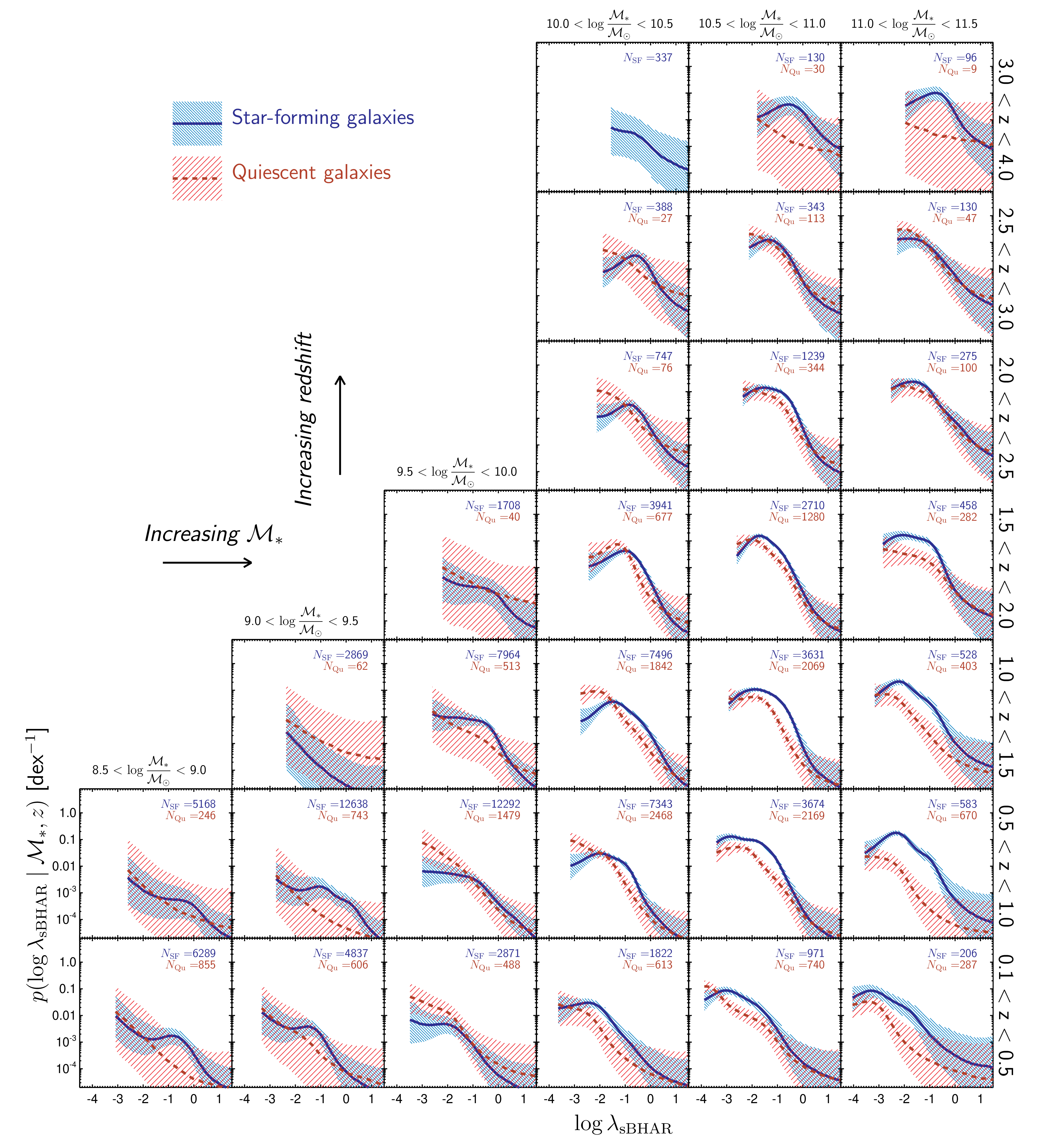}
\caption{
The probability distribution of specific black hole accretion rates, \Sar, within samples of star-forming galaxies (blue solid lines) and within samples of quiescent galaxies (red dashed lines), as a function of stellar mass and redshift.
Shaded regions indicate the 90 per cent confidence intervals on the recovered distributions. 
In each panel, we give the total number of star-forming galaxies ($N_\mathrm{SF}$) and quiescent galaxies ($N_\mathrm{Qu}$) within a stellar mass--redshift bin. 
}
\label{fig:pledd_sf_and_qu}
\end{figure*}

\begin{sidewaysfigure*}
\begin{center}
\vspace{0.8\textwidth}
\includegraphics[height=0.53\textwidth,trim=0 0 0 -20]{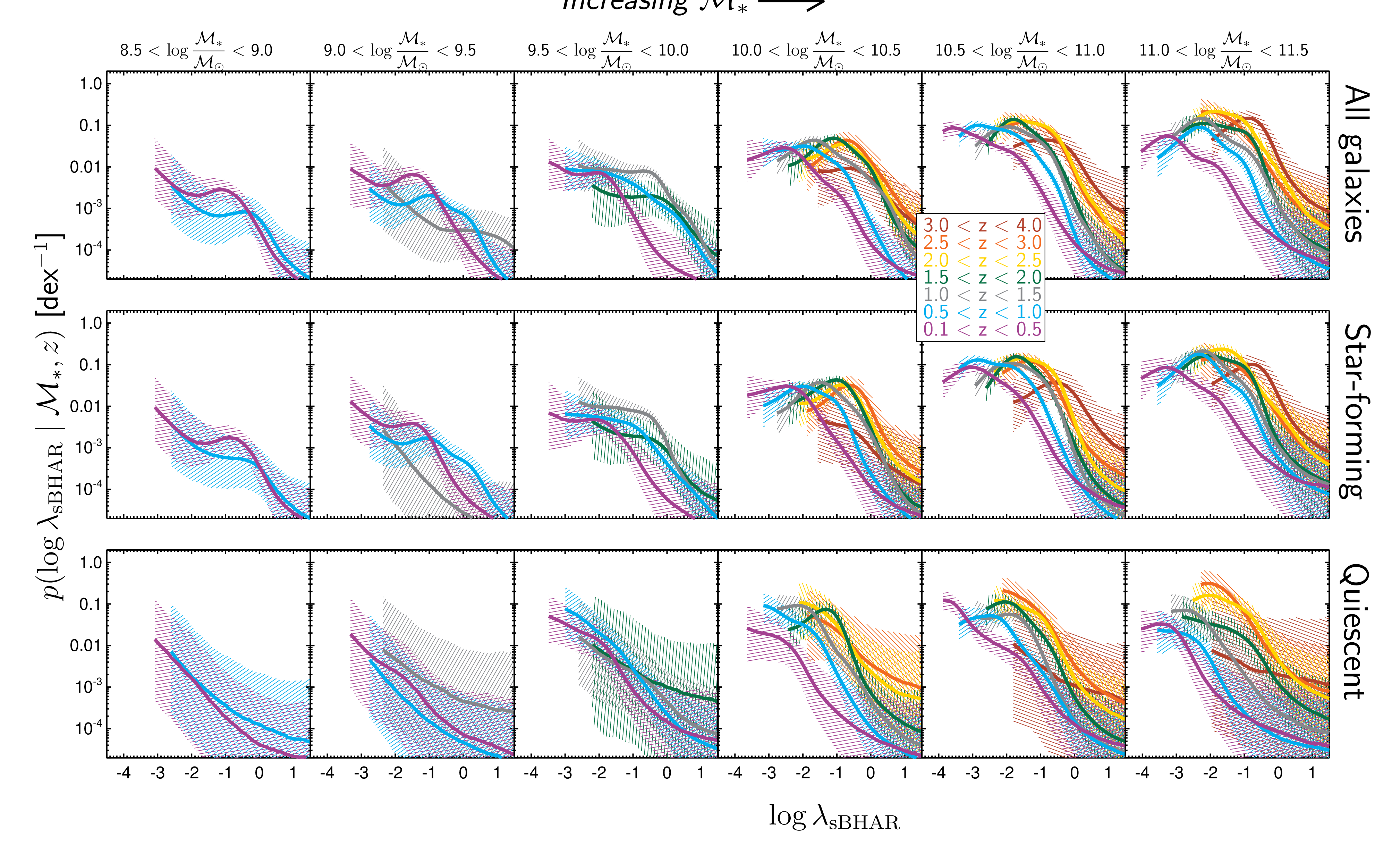}
\end{center}
\caption{
The probability distribution of specific black hole accretion rates, \Sar, as a function of redshift (as indicated by the different colours) in fixed stellar mass bins, for our full sample of all galaxies (top row), and dividing into the star-forming (middle row), and quiescent (bottom row) galaxy populations.
}
\label{fig:pledd_mbins}
\end{sidewaysfigure*}

\begin{sidewaysfigure*}
\begin{center}
\vspace{0.8\textwidth}
\includegraphics[height=0.53\textwidth,trim=0 0 0 -20]{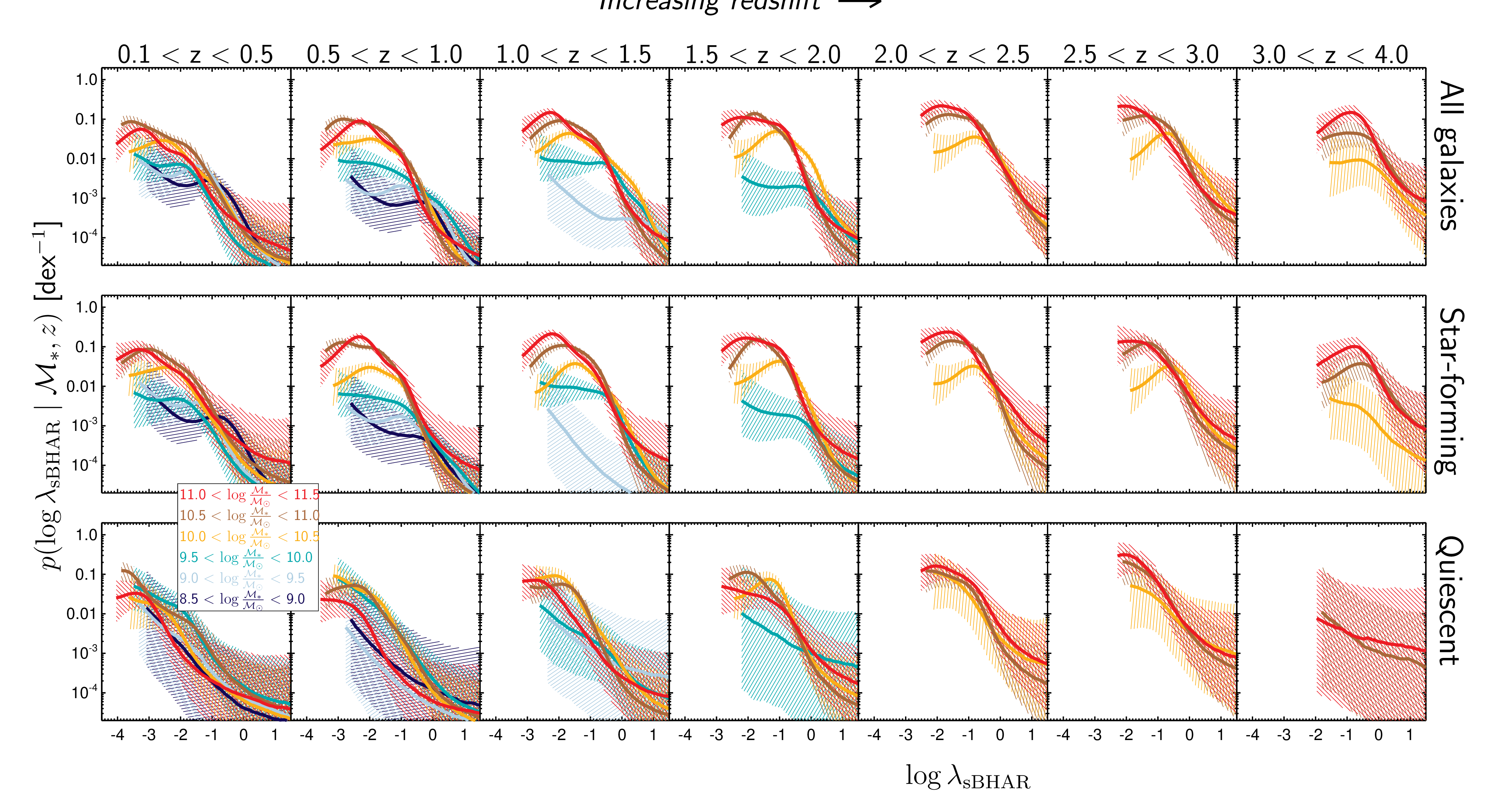}
\end{center}
\caption{
The probability distribution of specific black hole accretion rates, \Sar, as a function of stellar mass (as indicated by the different colours) in fixed redshift bins, for all galaxies (top row) and divided into the star-forming (middle row) and quiescent (bottom row) populations. 
We note that in higher redshift bins we lack measurements of \Psar\ at the lowest galaxy masses, due to the flux limits of our near-IR galaxy surveys. 
}
\label{fig:pledd_zbins}
\end{sidewaysfigure*}

\subsection{The distribution of accretion rates within the star-forming and quiescent galaxy populations}
\label{sec:sfqu}

The overall galaxy population is built up by a mixture of star-forming and quiescent galaxies that varies with stellar mass and redshift. 
The incidence and distribution of AGN activity could be very different within the different galaxy populations given their different gas content, morphologies and star formation histories that potentially complicate a physical interpretation of the overall distributions traced in Figure~\ref{fig:pledd_all}.
Thus, in Figure~\ref{fig:pledd_sf_and_qu} we divide our galaxy sample into star-forming and quiescent populations based on their SFRs (see Section~\ref{sec:data} for details) and present measurements of $p(\log \sar \giv \mstel,z)$ within each population, as a function of stellar mass and redshift.

We find AGNs with a wide range of \Sar\ in both star-forming and quiescent galaxies at all redshifts and stellar masses \citep[see also][]{Azadi15}.
In the majority of the stellar mass and redshift panels of Figure~\ref{fig:pledd_sf_and_qu}, we are unable to identify any significant differences in $p(\log \sar \giv \mstel,z)$ between the two galaxy populations, given the uncertainties. 
However, at higher stellar masses ($\log \mstel/\msun \gtrsim 10.0$) and lower redshifts ($z\lesssim2.0$), where we have the best constraints, there are significant differences whereby $p(\log \sar \giv \mstel,z)$ is higher at $\log \sar \gtrsim -2$ in star-forming galaxies.
Thus, the probability of finding a rapidly accreting AGN is higher by up to a factor $\sim$5 in a star-forming galaxy than in a quiescent galaxiy of equivalent stellar mass and redshift.\footnote{
We note that quiescent galaxies dominate the overall number density of galaxies at high stellar masses ($\log \mstel/\msun \gtrsim10.5$) and low redshifts ($z\lesssim 1$) and thus there are more quiescent galaxies in our samples, as indicated in Figure~\ref{fig:pledd_sf_and_qu}, whereas star-forming galaxies dominate at low masses and higher redshifts. 
Our measurements, however, track the \emph{fraction} of AGNs within a given galaxy sample. 
}
The turnover in $p(\log \sar \giv \mstel,z)$ at low \Sar\ is also seen more clearly for the star-forming galaxy samples.
The entire distribution may be shifted to higher \Sar\ for star-forming galaxies compared to the quiescent galaxies and $p(\log \sar \giv \mstel,z)$ may rise more steeply to lower \Sar\ in quiescent galaxies, although the uncertainties are large.

In Figure~\ref{fig:pledd_mbins} we examine the redshift evolution of \Psar\ at a given stellar mass within the entire galaxy population (top row), star-forming galaxies (middle row) and quiescent galaxies (bottom row), reproducing the results from Figures~\ref{fig:pledd_all} and \ref{fig:pledd_sf_and_qu} to reveal the dependence on redshift. 
There is no evidence for evolution in \Psar\ at low stellar masses ($\log \mstel/\msun < 10.0$) for any of the galaxy populations, although the uncertainties are large and we are only sensitive to a limited redshift range at these masses.
However, we observe strong evolution of \Psar\ in both star-forming and quiescent galaxies in all of the higher ($\log \mstel/\msun>10.0$) mass bins, which combine to produce the observed evolution for all galaxies.
The evolution appears to be characterised by an overall shift of \Psar\ to higher \Sar\ at higher redshifts for both star-forming and quiescent galaxies.
Interestingly, the evolution appears to be somewhat stronger in massive star-forming galaxies from $z\sim0.1-1$, with a rapid shift in the distribution toward higher \Sar, but with little or only mild evolution to higher redshifts.
In massive quiescent galaxies, evolution is seen over the entire redshift range, although the distributions are generally at lower \Sar, on average, compared to star-forming galaxies of equivalent mass and redshift.
We attempt to quantify these trends using estimates of the AGN duty cycle in Section~\ref{sec:dutycycle} below. 

Figure~\ref{fig:pledd_zbins} reproduces the results from Figures~\ref{fig:pledd_all} and \ref{fig:pledd_sf_and_qu} to reveal the stellar mass dependence in fixed redshift bins. 
For the quiescent galaxy population (bottom row of Figure~\ref{fig:pledd_zbins}) our constraints on \Psar\ are generally consistent with a single distribution with no stellar-mass dependence, although at lower redshifts ($z\lesssim1.5$) we find that \Psar\ may be shifted to lower \Sar\ at the highest stellar masses ($\mstel \gtrsim 10^{11}\msun$). 
Furthermore, our constraints on \Psar\ are poor at $\log \mstel/\msun \lesssim 10.0$ as our samples of quiescent galaxies at these masses are relatively small and thus we cannot rule out a stellar-mass dependence.
For star-forming galaxies (middle row of Figure~\ref{fig:pledd_zbins}), we find clear evidence of a stellar-mass dependence in \Psar\ at a fixed redshift, at least at $0.5\lesssim z \lesssim 2$ where we have good constraints. 
At a given redshift, the maximum accretion rate appears to be limited by the same steep power law, while 
at lower accretion rates the distribution has a mass-dependent normalisation such that the probability of a star-forming galaxy hosting an AGN with $-3 \lesssim \log \sar \lesssim -1$ is suppressed at lower stellar masses. 
The total galaxy population (top row of Figure~\ref{fig:pledd_zbins}) also shows evidence for a stellar-mass dependence at a fixed redshift (to at least $z\sim2$).
The mixing of the star-forming and quiescent populations leads to the complex structure and mass-dependence seen in \Psar\ for the overall galaxy sample that is difficult to interpret without considering the constituent populations separately.

\subsection{The AGN duty cycle and the average AGN accretion rate}
\label{sec:dutycycle}

\begin{figure*}
\includegraphics[width=\textwidth,trim=20 20 20 0]{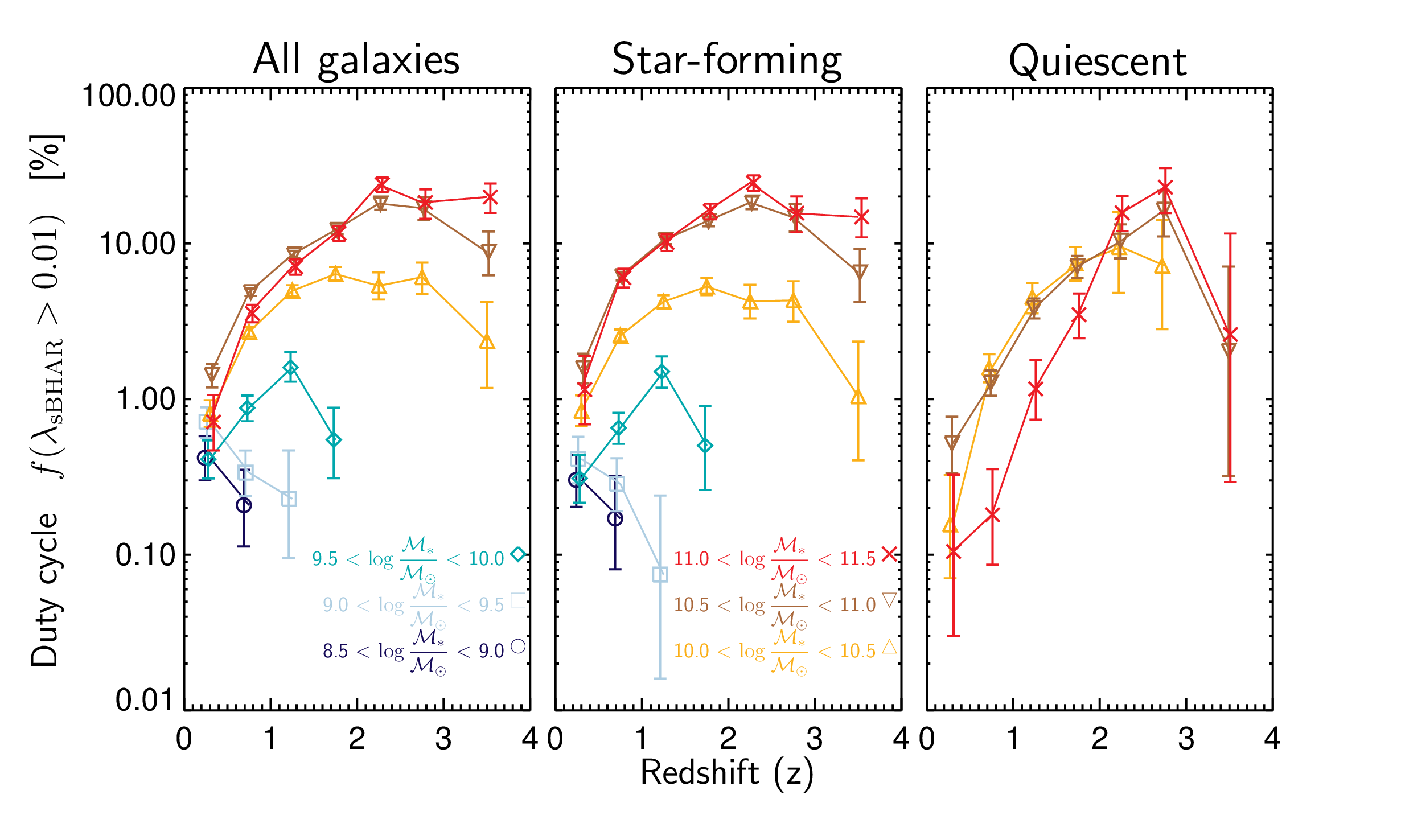}
\caption{
AGN duty cycle (the fraction of galaxies with black holes accreting at $\sar>0.01$) as a function of redshift at different stellar masses (as indicated by the colours and symbols) for all galaxies (left), star-forming galaxies (centre) and quiescent galaxies (right). 
The duty cycle is calculated by integrating our estimates of \Psar\ from Section~\ref{sec:accdists} (see Equation~\ref{eq:dutycycle}). 
Error bars indicate $1\sigma$-equivalent confidence intervals, based on the posterior distributions of \Psar\ from our Bayesian analysis. 
For quiescent galaxies we only show estimates for our three highest mass bins where we have reasonable constraints on the duty cycle.
}
\label{fig:dutycycle_vs_z}
\end{figure*}

\begin{figure*}
\includegraphics[width=\textwidth,trim=20 30 20 0]{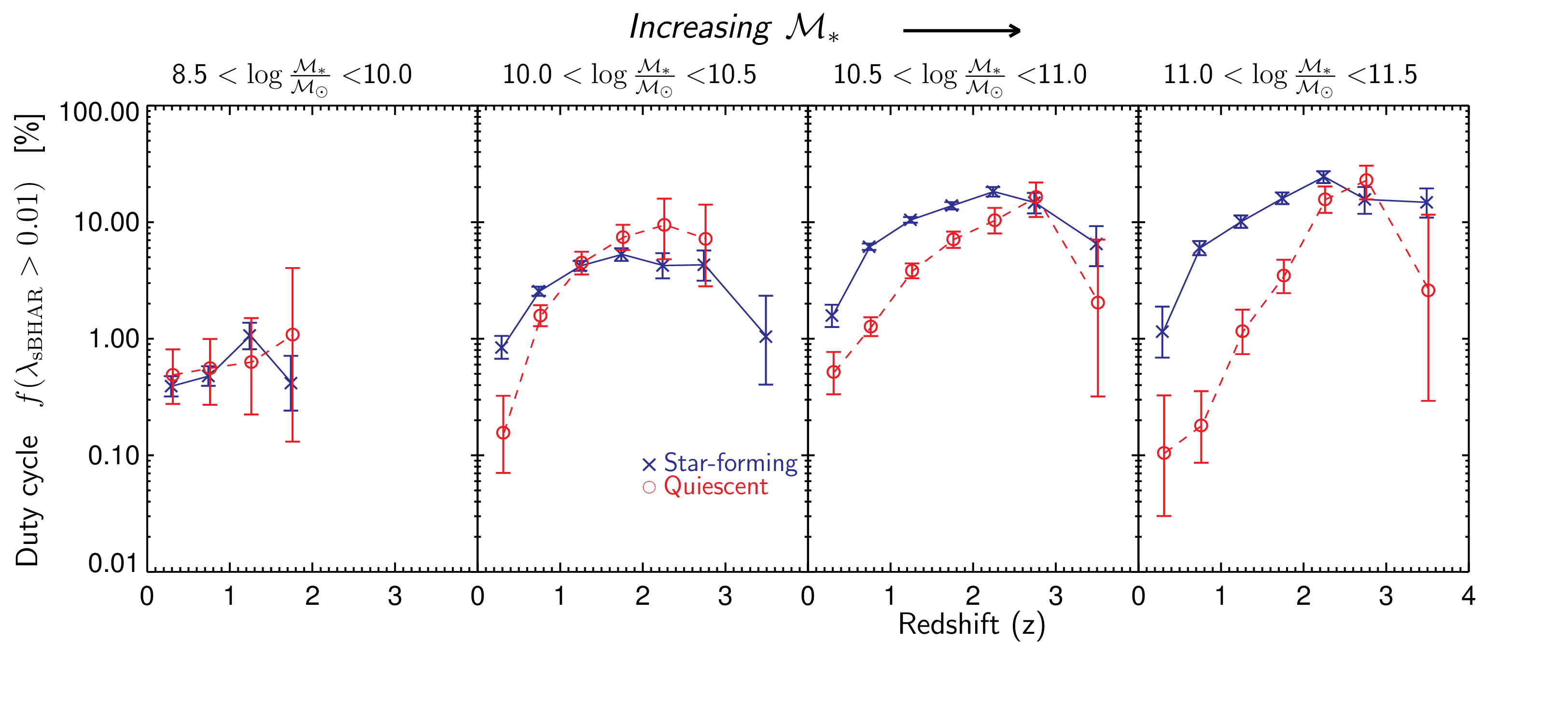}
\caption{
AGN duty cycle as a function of redshift at different stellar masses within the star-forming (blue crosses) and quiescent  
(red circles) galaxy populations. 
We note that the lowest stellar mass bin (leftmost panel) includes galaxies with $8.5<\log \mstel/\msun<10.0$ and above our stellar mass completeness limits (and thus corresponds to a higher average stellar mass at higher redshifts). 
We find that the duty cycle increases with redshift within both star-forming and quiescent galaxies with $\mstel\gtrsim10^{10}\msun$, indicating that a higher proportion of galaxies are rapidly growing their black holes at higher redshifts. 
For massive galaxies ($\mstel>10^{10.5}\msun$) the duty cycle for quiescent galaxies is lower than for star-forming galaxies at $z\lesssim2$ but evolves more rapidly with redshift.
}
\label{fig:dutycycle_vs_z_massbins}
\end{figure*}

\begin{figure*}
\includegraphics[width=\textwidth,trim=20 20 20 0]{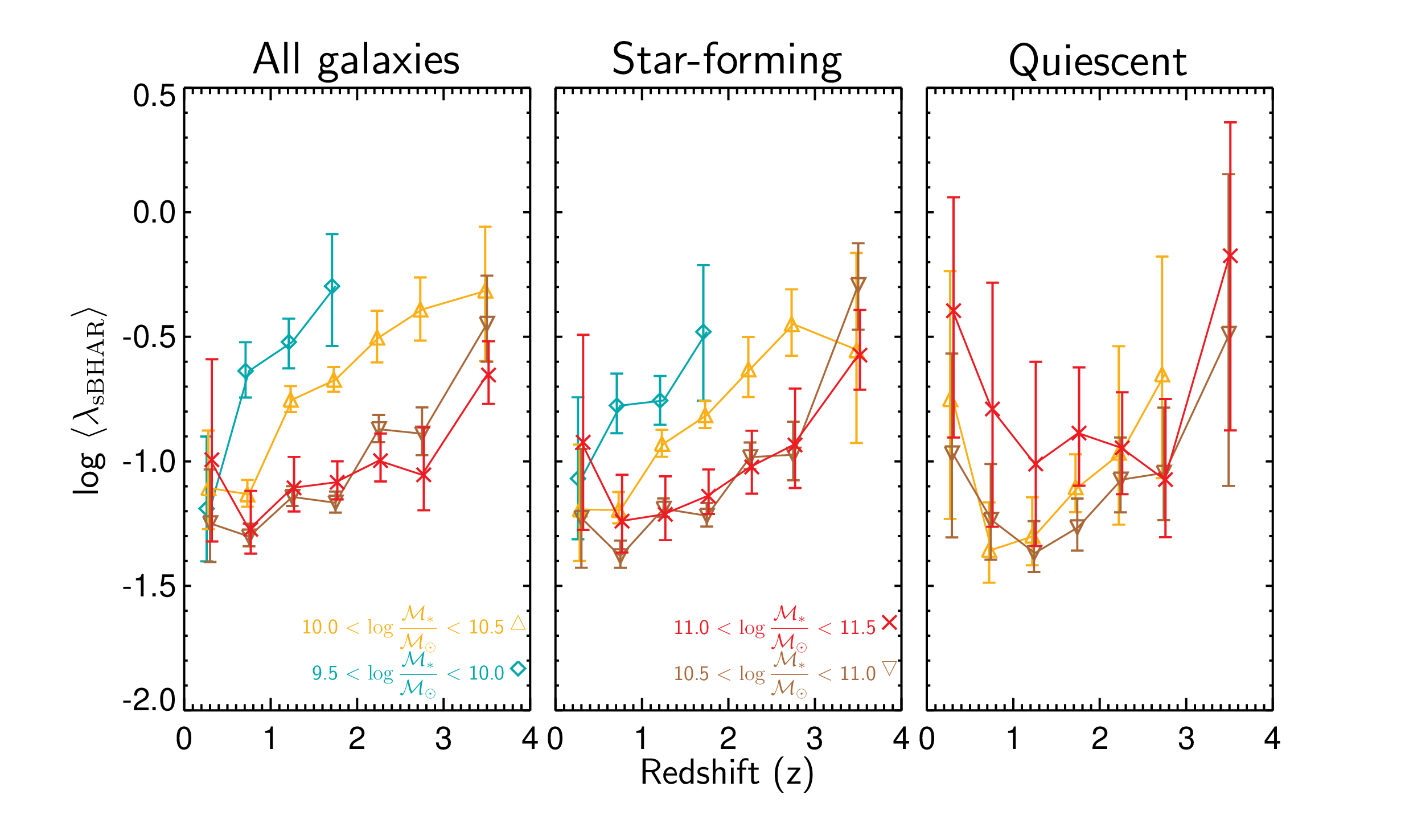}
\caption{
\upd{
Average accretion rate, $\langle \sar \rangle$, for AGNs accreting above a limit of $\sar>0.01$ (calculated using Equation~\ref{eq:avl}) as a function of redshift and at different stellar masses (as indicated by the colours and symbols) for all galaxies (left), star-forming galaxies (centre) and quiescent galaxies (right). 
Error bars indicate 1$\sigma$-equivalent confidence intervals.
For clarity, we omit the estimates in the two lowest stellar mass bins ($8.5<\log \mstel/\msun<9.0$ and $9.0<\log \mstel/\msun<9.5$) for all galaxies and star-forming galaxies due to the large uncertainties, although we note that the estimates are consistent with the general trend of increasing $\langle \sar \rangle$ toward lower stellar masses (at fixed redshift). 
As in Figure~\ref{fig:dutycycle_vs_z}, we omit the lowest stellar mass bins for the quiescent galaxies, where both the duty cycle and $\langle \sar \rangle$ are very poorly constrained due to the small sample sizes.
}
}
\label{fig:avl_vs_z}
\end{figure*}

In this section we attempt to summarize and quantify the information contained in Figures~\ref{fig:pledd_all}--\ref{fig:pledd_zbins} by calculating the duty cycle of AGN activity as a function of stellar mass \upd{as well as the average accretion rate associated with these AGNs.}
\upd{We are thus able to condense the information contained in the full accretion rate distributions down to just two numbers, making it easier to identify and interpret the overall trends with redshift, stellar mass or galaxy type, albeit at the loss of some information on the overall shapes of the distributions as a function of \Sar.}

We define the AGN duty cycle as the fraction of galaxies (at a given range of stellar mass and redshift) with supermassive black holes growing at $\sar > 0.01$. 
We can calculate the AGN duty cycle from our measurements of \Psar\ described above,
\begin{equation}
f(\sar > 0.01) = \int_{-2}^\infty p(\log \sar \giv \mstel,z) \; d\log \sar 
\label{eq:dutycycle}
\end{equation}
where $f(\sar > 0.01)$ corresponds to our definition of the duty cycle. 
Uncertainties in the duty cycle are propagated from the posterior distributions of \Psar\ and correspond to the 1$\sigma$-equivalent (i.e. 68~per cent) central confidence intervals. 

Figure~\ref{fig:dutycycle_vs_z} presents measurements of the duty cycle, $f(\sar>0.01)$, as a function of redshift for different stellar masses.
A number of trends, previously identified in the full accretion distributions in Sections~\ref{sec:all} and \ref{sec:sfqu} above, are apparent in this figure. 
Within star-forming galaxies, we observe a clear stellar mass dependence between $\log \mstel/\msun\approx10.0$ and $\log \mstel/\msun\approx 11.5$, whereby the duty cycle is higher in higher stellar mass galaxies at any given redshift. 
At high stellar masses ($\log \mstel/\msun \gtrsim 10.0$), we find that the duty cycle evolves with redshift, increasing up to $z\sim2$ and flattening off or potentially declining to higher redshifts. 
The redshift evolution appears to get weaker at lower stellar masses and in our lowest stellar mass bins ($\log \mstel/\msun <9.5$) our results are consistent with no evolution (although we only probe a limited range of redshifts and the uncertainties are large). 
For quiescent galaxies, conversely, the duty cycle is generally consistent with \emph{no} stellar mass dependence between $\log \mstel/\msun \approx10$ and 
$\log \mstel/\msun \approx 11$ and a drop at the highest stellar mass ($11<\log \mstel/\msun<11.5$), at least out to $z\sim2$.
The data for quiescent galaxies are consistent with a strong evolution with redshift for all stellar masses between $\log \mstel/\msun=10.0$ and $\log \mstel/\msun=11.5$; we do not show data for the lower stellar mass bins where there are very few quiescent galaxies and the duty cycle is poorly constrained.

In Figure~\ref{fig:dutycycle_vs_z_massbins} we directly compare the redshift evolution of the duty cycle in star-forming and quiescent galaxies at a range of stellar masses. 
For the lowest mass bin we include all galaxies (of a given type) with $8.5<\log \mstel/\msun<10.0$ \emph{and} above our stellar mass completeness limits. 
At these low masses there is no evidence of a difference in the duty cycle between star-forming or quiescent galaxies or any evolution with redshift
(although we cannot rule out an increase of the duty cycle with redshift for quiescent galaxies, given the large uncertainties). At $10.0<\log \mstel/\msun<10.5$ we find that the duty cycle increases between $z\sim0$ and $z\sim1-2$ for both galaxy types and is roughly constant to higher redshifts.
The duty cycle is consistent between the the two galaxy types at $z>1.0$ (within the uncertainties) but at lower redshifts we measure a significantly lower duty cycle (by a factor $\sim 2-6$) in quiescent galaxies compared to star-forming galaxies. 
In the two highest mass bins in Figure~\ref{fig:dutycycle_vs_z_massbins} we find that the duty cycle is substantially lower in quiescent galaxies at $z<2$ (by a factor $\sim 10$) compared to star-forming galaxies but evolves more rapidly with redshift.
The full distributions shown in Figure~\ref{fig:pledd_sf_and_qu} confirm this trend. 
At $z>2$ the duty cycles in star-forming and quiescent galaxies are fairly similar in the two high mass bins, although we note that at these redshifts there are substantially fewer quiescent galaxies than star-forming galaxies (even at these high stellar masses). Thus, the bulk of AGN at high redshifts are hosted by star-forming galaxies, despite the similar duty cycles in both galaxy types.

\upd{To supplement the AGN duty cycle and gain further insights into the accretion rates of the AGNs contained within our galaxy samples, we calculate the average accretion rate, $\langle \sar \rangle$, for galaxies that contain AGNs accreting above a fixed limit of $\sar>0.01$, given by}
\begin{equation}
\langle \sar \rangle = 
\frac{\displaystyle \int_{-2}^{\infty} p(\log \sar \giv \mstel,z) \; \sar \; d\log \sar } 
     {\displaystyle \int_{-2}^{\infty} p(\log \sar \giv \mstel,z) \; d\log \sar }  .
\label{eq:avl}
\end{equation}
\upd{While the duty cycle represents the \emph{fraction} of galaxies with AGNs, the average accretion rate defined in this way provides information on the \emph{typical} accretion rate of \emph{those} AGNs, where in both cases we are effectively defining an ``AGN" as a supermassive black hole that is accreting above a rate of $\sar=0.01$.
This quantity thus allows us to distinguish between galaxies with similar overall duty cycles but where the shapes of \Psar (above the $\sar=0.01$ limit) are very different, e.g. where the bulk of those AGNs are close to the limit versus at much higher \Sar.}

\upd{
Our estimates of $\langle \sar \rangle$ are presented in Figure~\ref{fig:avl_vs_z}. 
Our measurements reveal a clear pattern within both star-forming galaxy and all galaxy samples (dominated by star-forming galaxies at all stellar masses for $z\gtrsim1$), 
whereby the average accretion rate is increasing with redshift at all stellar masses (increasing by $\sim0.5$~dex between $z\sim0.5$ and $z\sim3$) and becomes progressively \emph{higher} with decreasing stellar mass at a fixed redshift.}

\begin{figure*}
\includegraphics[width=\textwidth,trim=0 30 0 10]{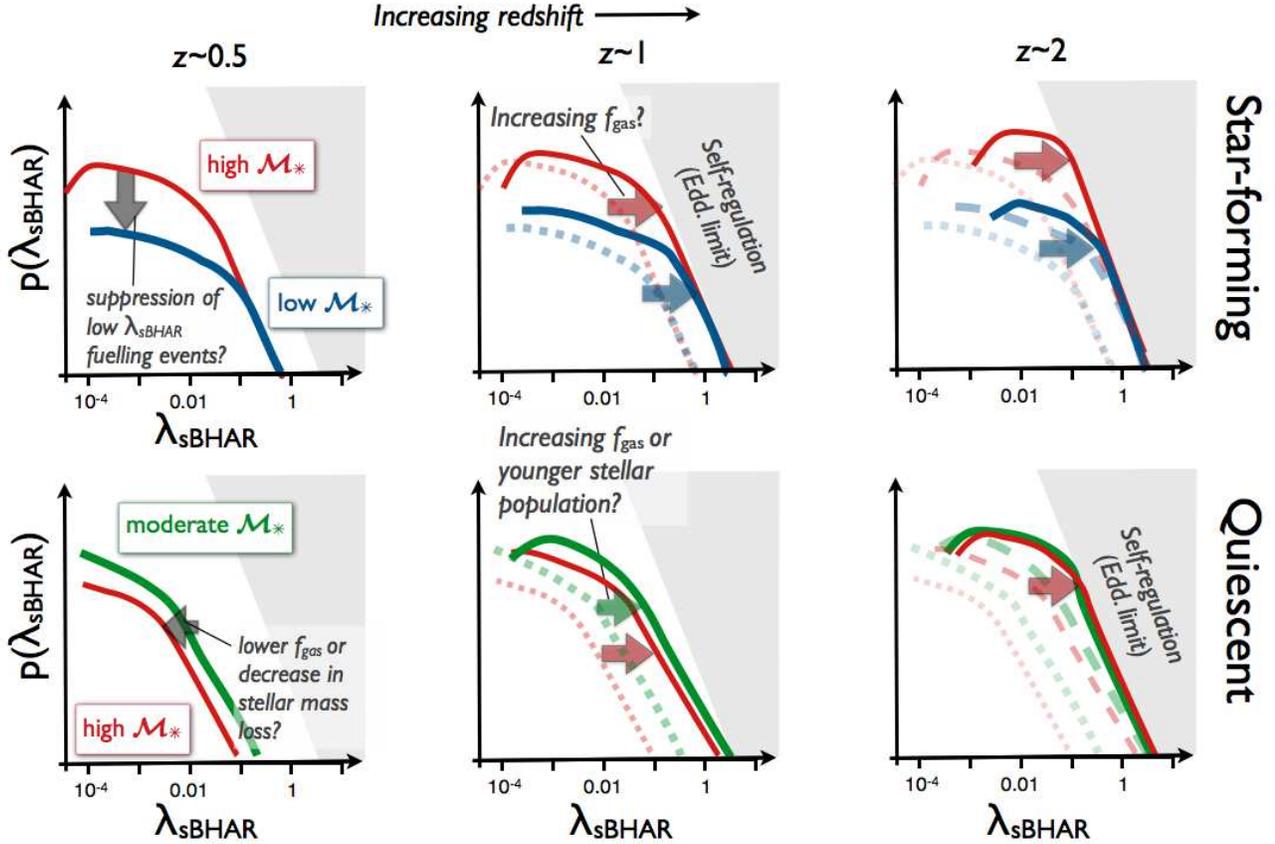}
\caption{
Sketch summarizing the broad trends identified in this paper for the shape, redshift evolution and stellar mass dependence of the probability distribution of specific black hole accretion rates within galaxies of given stellar mass and redshift, $p(\log \sar \giv \mstel,z)$ -- for simplicity denoted as $p(\sar)$ here -- considering both star-forming galaxies (top row) and quiescent galaxies (bottom row). See Section \ref{sec:discuss} for a full discussion. 
}
\label{fig:sketch}
\end{figure*}

This behaviour contrasts with the evolution of the duty cycle (see Figure~\ref{fig:dutycycle_vs_z}), where the duty cycle becomes progressively \emph{lower} with decreasing stellar mass. 
Thus, while the fraction of galaxies hosting an AGN is lower in lower massive galaxies, the typical accretion rates of that small fraction of AGNs are generally higher.
The underlying cause of this pattern is more apparent in the full accretion rate distributions (e.g. central row of Figure~\ref{fig:pledd_zbins}), where we can see at a fixed redshift (e.g. $0.5<z<1.0$) that \Psar\ is suppressed at lower accretion rates in lower mass galaxies (reducing the overall duty cycle) but that the distribution extends to higher values (the break in the distribution is at higher \Sar), resulting in the increase in $\langle \sar \rangle$. 
In the discussion below (Section \ref{sec:discuss}) we present a simplified sketch that attempts to summarize these overall trends (see Figure~\ref{fig:sketch}).

\upd{
Our measurements of $\langle \sar \rangle$ within quiescent galaxies (right panel of Figure~\ref{fig:avl_vs_z}) are fairly noisy, although at $10<\log \mstel/\msun<10.5$ and $10.5<\log \mstel/\msun < 11$ (yellow and brown points) we do see a rise of the average accretion rate between $z\sim0.5$ and $=z\sim3$, similar to that seen in the star-forming galaxy sample. 
We note that the average accretion rates are generally lower for the quiescent galaxies than for star-forming galaxies at the same redshift and stellar mass, consistent with the differences seen in Figure~\ref{fig:pledd_sf_and_qu}.
We also caution that our measurements are most uncertain in our lowest redshift bin ($0.1<z<0.5$) for all galaxy types, where the duty cycle is generally lowest and it is hard to identify clear trends.
}

\upd{
In summary, our measurements show clear evidence that the duty cycle of AGN, defined above a fixed \Sar\ threshold,  increases with stellar mass in star-forming galaxies,
while the typical accretion rates of these AGNs are lower at higher stellar masses.
The duty cycle in quiescent galaxies is generally lower than in star-forming galaxies and decreases at the highest stellar masses ($11<\log \mstel/\msun<11.5$ at $z\lesssim2$).
Both the duty cycle and the average accretion rates increase with redshift in star-forming and quiescent galaxies.
The strongest evolution of the duty cycle is observed in high stellar mass quiescent galaxies ($11<\log \mstel/\msun<11.5$), where the duty cycle increases from $f(\sar>0.01)\approx0.1$ per cent at $z\approx0.3$ to $f(\sar>0.01)\approx10$ per cent at $z\approx2$. 
In Figure~\ref{fig:sketch} below we attempt to encapsulate all these trends into a simplified sketch summarizing our measurements of the full accretion rate probability distributions, \Psar, as a function of stellar mass and redshift in both the star-forming and quiescent galaxy populations (see Section~\ref{sec:discuss} for a full discussion). 
}

\section{Discussion}
\label{sec:discuss}

In this paper we have presented new measurements of the probability distribution function of AGN accretion rates, \Psar\footnote{For simplicity, we denote \Psar\ as $p(\sar)$ throughout the remainder of this discussion section, implicitly assuming logarithmic units and that the probability distribution is defined within a sample of galaxies of given \Mstel\ and $z$.},
\renewcommand{\Psar}{$p(\sar)$} 
as a function of stellar mass and redshift within both the star-forming and quiescent galaxy populations. 
Our method (developed in the companion \PaperI) starts with near-infrared selected samples of galaxies and extracts X-ray information for \emph{every} galaxy from deep \textit{Chandra} imaging. 
We are thus able to measure the \emph{probability} of hosting an AGN directly (as a function of \Sar) for different galaxy types and place robust constraints on the shape of the accretion rate distribution (using a non-parametric model with minimal assumptions) over a wide range in both stellar mass and redshift. 
Our measurements reveal a somewhat complex picture where \Psar\ differs between galaxy types and depends on both stellar mass and redshift. 
Here, we present a simplified summary of our findings, discuss the possible physical underpinnings of the observed behaviour, and compare our measurements with previous studies.

\subsection{The distribution and evolution of AGN activity in star-forming galaxies} 
\label{sec:discuss_sf}

In the top row of Figure \ref{fig:sketch} we attempt to summarize our main findings for the shape, redshift evolution and stellar mass dependence of \Psar\ in star-forming galaxies. 
We note that this figure is intended as an illustration to guide the discussion and interpretation of our results, as recovered by our non-parametric method. 
A full parametric modelling of our data, informed by these findings, is deferred to future work. 

In low-redshift ($z\sim0.5$) star-forming galaxies (illustrated in the top-left panel of Figure~\ref{fig:sketch}) our results are consistent with a broad distribution of accretion rates with a relatively shallow slope at lower accretion rates (with exponent $\sim0.2-0.8$, consistent with prior findings e.g. \citetalias{Aird12}; \citealt{Bongiorno12,Azadi15}), a break at $\sar \approx 0.1$, 
and a steeper power law extending to higher redshifts (as inferred by \citealt{Aird13,Caplar15,Bongiorno16}; see also \citealt{Georgakakis17}).
This broad distribution likely reflects long-term variability in the levels of AGN accretion within a given star-forming galaxy over timescales $\lesssim100$~Myr \citep{Hickox14,Schawinski15} due to the stochastic nature of 
the supply of gas to the nuclear regions of galaxies, 
in addition to localised heating and feedback from either the AGN itself or stellar processes \citep[see e.g.][]{Hopkins06c,Novak11,Gaspari17}. 
At these redshifts ($z\sim0.5$), the break at $\sar\approx 0.1$  indicates a lack of high accretion rate events, possibly related to the distribution of cold molecular gas within low-redshift star-forming galaxies (and a lack of massive gas clouds) or the rarity of violent triggering events (e.g. mergers) that can drive sufficiently high gas densities into the central regions to fuel the highest accretion rate AGN \citep[e.g.][]{Kauffmann00,Hopkins08}

We also identify a \refone{possible} turnover or flattening of \Psar\ at the lowest accretion rates ($\sar \lesssim 10^{-2}-10^{-3}$), \refone{at least at moderate-to-high stellar masses ($\log \mstel/\msun \gtrsim10$) where we are able to probe sufficiently low accretion rates out to $z\sim2$.}
\refone{
We note that the position of the low-\Sar\ turnover is close to the regime where star formation processes dominate (see e.g. Figure~\ref{fig:pledd_all}), making it difficult to robustly determine the full extent of any turnover.}
An eventual flattening or turnover at low accretion rates is not surprising. 
Indeed, a flattening is \emph{required} to ensure that the probability distribution function integrates to less than unity -- a reasonable prior that is implicit in our Bayesian modelling.
However, the position of the observed turnover is not driven by this prior as the integral of \Psar\ is generally much less than 1 over the range of \Sar\ that we probe (i.e. we do not reach such low accretion rates that \emph{every} galaxy effectively hosts an AGN).
The low-\Sar\ turnover may therefore represent a natural lower limit to the range of accretion rates in galaxies, indicating that a minimal fuelling level produces a short-lived radiatively efficient AGN (observable in the X-ray) that consumes the gas in the central region and rapidly fades, with the black hole then remaining in a state of relative quiescence for a period of time\footnote{Very low accretion rate, radiatively inefficient AGN activity, seen in the radio, may be present during this time but is not revealed by our X-ray--based measurements.} before a subsequent fuelling event triggers the next AGN episode.

Our measurements also reveal a stellar mass dependence whereby the probability of a lower mass galaxy hosting a low-accretion-rate AGN ($\sar \lesssim 0.1$) is suppressed compared to the probability within higher mass galaxies (as indicated in Figure~\ref{fig:sketch}), although a broad range of accretion rates---spanning the range observed in higher mass galaxies---are still observed. 
This suppression of low \Sar\ fuelling events could indicate that it is more difficult to drive smaller pockets of gas into the central regions of lower mass galaxies, possibly due to the difficulty of releasing angular momentum from such gas clouds in lower mass galaxies \citep[e.g.][]{Rosas-Guevara15} or an increase in the effects of stellar  feedback at low galaxy masses \citep[e.g.][]{Dekel86,White91,Cole00,Hopkins14}. 
The nature of this low-\Sar\ suppression reduces the overall \emph{probability} of hosting an AGN at lower \Sar\ (resulting in a lower duty cycle in low-mass star-forming galaxies, see Figure~\ref{fig:dutycycle_vs_z}), yet the observed \emph{range} of accretion rates remains the same and there is thus an increase in the average accretion rate 
(see Figure~\ref{fig:avl_vs_z}).

As redshift increases (from $z\sim0.5$ to $z\sim1$), the accretion rate distributions in star-forming galaxies appear to shift systematically toward higher \Sar\ (as illustrated in Figure~\ref{fig:sketch}; see also the measurements in Figure~\ref{fig:pledd_mbins}). 
We propose that this shift is related to the substantial increase in the cold gas fraction toward higher redshifts within star-forming galaxies at all stellar masses.
Such an increase is also thought to drive the evolution of the main sequence of star formation and the associated increase in the typical SFRs of star-forming galaxies to higher redshifts \citep[e.g.][]{Lagos11,Combes13,Saintonge13,Santini14}. 
This evolution of \Psar\ toward higher \Sar\ drives the rapid evolution of the duty cycle (above our $\sar>0.01$ threshold) in higher mass galaxies due to the transition from the steeper high-\Sar\ slope to the flatter low-\Sar\ slope in this regime.
The average accretion rate, $\langle \sar \rangle$, on the other hand, increases in a consistent manner across all stellar masses.

Our results indicate that the accretion rate distributions are ultimately limited by the self-regulation of black hole growth, at a point that roughly corresponds to the Eddington limit ($\sar\approx1$ in our adopted units), indicated by the grey shaded region in Figure~\ref{fig:sketch}. 
Rather than a hard limit, however, the data suggest a steep power-law relation defines this limit as rare and short-lived periods of super-Eddington accretion may still take place and, more importantly, scatter in the black hole mass-to-stellar mass scaling relation will soften a hard cut-off in our observations of \Sar\ (we do not directly measure the Eddington ratio, see \citealt{Aird13}). 
Effectively, galaxies with over-massive black holes (compared to the baseline relation) can grow at specific accretion rates (relative to the host mass) that exceed $\sar\approx1$ without violating the true Eddington limit of the black hole.
\updd{Such specific accretion rates ($\sar>1$) would also be observed if black hole growth precedes the bulk of galaxy growth 
\citep[e.g.][]{Bennert11,Volonteri15,Trakhtenbrot15}.}

We suggest that the self-regulation of black hole growth prevents further positive evolution of \Psar\ at high \Sar\ above $z\sim1$ in star-forming galaxies, despite further increases in the gas fraction.
While more gas is supplied to the central regions of the galaxy, self-regulation of the accretion process prevents \Psar\ from exceeding this limit. 
Nevertheless, the copious supply of gas may continue to push up the average accretion rates (see Figure~\ref{fig:avl_vs_z}), increase the \emph{minumum} accretion rates (i.e. the low-\Sar\ turnover) and lead to a narrowing of the distribution close to the self-regulation limit at the highest redshifts (as illustrated in the top-right panel of Figure~\ref{fig:sketch}).

\subsection{The distribution and evolution of AGN activity in quiescent galaxies}
\label{sec:discuss_qu}

We now turn our attention to the AGNs within quiescent galaxies. 
The bottom row of Figure~\ref{fig:sketch} attempts to summarize our main results and the overall trends that we identify with redshift and stellar mass, focusing on moderate-to-high stellar masses ($\log \mstel/\msun\sim10-11.5$), where our measurements of \Psar\ are most reliable. 

As in star-forming galaxies, we measure a broad distribution of \Sar\ within quiescent galaxies across all stellar masses and redshifts that we probe, likely indicating that stochastic fuelling mechanisms are at work. 
At lower redshifts (e.g. $z\sim0.5-1$), the distribution of \Sar\ in quiescent galaxies is shifted to lower values than in star-forming galaxies, possibly reflecting the much lower gas fraction in the quiescent galaxy population. 
There are also indications of a mild stellar mass dependence at $z<2$, whereby at the highest stellar masses ($11<\log \mstel/\msun <11.5$) the duty cycle \emph{decreases} (see right panel of Figure~\ref{fig:dutycycle_vs_z}) and the accretion distribution may be shifted to lower \Sar\ on average (see bottom row of Figure~\ref{fig:pledd_zbins} and the illustration in Figure~\ref{fig:sketch}). 
Such a shift to lower \Sar\ could be due to a decrease in the gas fraction in the highest mass quiescent galaxies \citep[e.g.][]{Sargent15}, \upd{potentially due to the denser environments of higher mass quiescent galaxies and an associated decrease in the external gas supply.} 
Alternatively, the primary fuel supply for AGN activity in quiescent galaxies 
may be mass-loss from the stars in the galaxy itself \citep[e.g.][]{Norman88,Ciotti07,Kauffmann09}, which can provide a sustained supply of low angular momentum gas that could drive low-level AGN activity.
The shift to lower \Sar\ at the highest stellar masses would then reflect the older stellar ages of the most massive galaxies and the decreased mass loss rate from a more evolved stellar population.

Our measurements reveal a strong evolution of \Psar\ within quiescent galaxies for all masses $\log \mstel/\msun \gtrsim 10$  (see bottom row of Figure~\ref{fig:pledd_mbins}), which is reflected by the rapid increase in the AGN duty cycle (see right panel of Figure~\ref{fig:dutycycle_vs_z}). 
As in the star-forming galaxy population, this evolution may simply be driven by the overall increase in the gas fraction; at $z\gtrsim1$, the ``quiescent" galaxy population, defined relative to the evolving star-forming main sequence, can have relatively high SFRs (compared to quiescent galaxies at lower $z$) and may have correspondingly higher gas fractions \citep[e.g.][]{Gobat17}. 
Alternatively, the evolution may reflect the younger stellar ages of some higher redshift quiescent galaxies and the associated increase in stellar mass loss. 
At high redshifts ($z\sim2$), a significant fraction of massive quiescent galaxies have relatively young stellar populations, indicating they have undergone recent and rapid quenching of their star formation activity
\citep[e.g.][]{Whitaker12b}, 
which could explain the rapid evolution of \Psar\ from $z\sim0.5$ to 2 and the high duty cycle in quiescent galaxies at $z\gtrsim2$ \citep[see also][]{Wang17}. 

We also suggest that AGN activity in quiescent galaxies is subject to the same self-regulation---associated with Eddington-limited growth---that restricts the evolution of \Psar\ in star-forming galaxies. 
However, quiescent galaxies appear to reach this limit at higher redshifts ($z\gtrsim2$, compared to $z\sim1$ for star-forming galaxies), with high \Mstel\ quiescent galaxies reaching this limit at higher $z$ than in more moderate \Mstel\ galaxies (as illustrated in Figure~\ref{fig:sketch}).

\subsection{Comparison with prior studies of AGN accretion rate distributions}
Here we compare our findings with prior studies of the distribution of AGN accretion rates (or Eddington ratios) and their relation to the galaxy population.

Early work by \citet{Kauffmann09} used optical spectroscopic diagnostics to determine the distribution of Eddington ratios within nearby ($z\lesssim0.3$) galaxies from the Sloan Digital Sky Survey \citep[SDSS:][]{York00} and identified two, distinct Eddington ratio distributions: a relatively narrow, lognormal distribution in young (i.e. star-forming) galaxies; and a broader, power-law distribution in older (i.e. quiescent) galaxies. 
However, more recent work by \citet{Jones16} showed that the intrinsic distribution of Eddington ratios in star-forming galaxies---once contamination of the optical emission lines due to star formation processes is accounted for---is much broader and can be described by a Schechter function (i.e. a power-law like dependence at low Eddington ratios with an exponential cutoff at the highest accretion rates). 
Our measurements of \Psar\ based on the X-ray emission
thus appear consistent with lower redshift optical studies: 
we also find a broad distribution of \Sar\ in both star-forming and quiescent galaxies with a relatively flat, power-law like slope at lower \Sar\ and a steeper tail to the highest \Sar.
In addition, we identify a turnover or flattening of \Psar\ at the lowest accretion rates ($\sar\lesssim 10^{-2} - 10^{-3}$) that may indicate a rapid consumption of gas in the nuclear regions by the central black hole, somewhat akin to the ``feast" mode introduced by \citet{Kauffmann09}, although our distributions are broader than the lognormal originally identified by those authors and are in better agreement (at higher \Sar) with the Schechter function proposed by \citet{Jones16}. 
Nevertheless, the differences between \Psar\ for star-forming and quiescent galaxies that we identify are indicative of different modes of black hole growth in the different galaxy types and are thus consistent with the general picture of a ``feast" mode (associated with the rapid consumption of cold gas) dominating in star-forming galaxies and a ``famine" mode (fuelled by stellar mass loss) in quiescent galaxies, as proposed by \citet{Kauffmann09}.

Prior work at higher redshifts have primarily used X-ray emission as the tracer of the AGN accretion rate and are thus more directly comparable with our study. 
\citetalias{Aird12} measured \Psar\ out to $z\sim1$ based on X-ray selection of AGNs within a sample of galaxies from the PRIMUS survey, and concluded that the distribution of accretion rates could be described by a \emph{mass-independent}, power-law distribution \citep[see also][]{Bongiorno12}.
In contrast to \citetalias{Aird12}, our new results presented here find evidence for a stellar-mass-dependent \Psar, both within the entire galaxy population and the star-forming and quiescent galaxy samples (see Figures~\ref{fig:pledd_all} and \ref{fig:pledd_sf_and_qu}). 
\updd{However, it is important to note our findings are consistent (within the uncertainties) with these prior studies.} 
In fact, the strong, observational bias identified by \citetalias{Aird12}---whereby lower specific accretion rate AGNs are harder to identify in lower mass galaxies---remains a significant effect in our study. 
By measuring the distribution of \emph{specific} accretion rates, normalizing by the stellar mass, we account for the observational bias identified by \citetalias{Aird12}. 
The stellar mass dependencies revealed by our measurements corresponds to a second-order effect that is in addition to the primary \emph{observational} effect identified by \citetalias{Aird12}. 
Our order of magnitude increase in galaxy sample size, the increase in the range of stellar masses and redshifts probed by our study, and our advanced techniques to exploit the deep \emph{Chandra} imaging have allowed us to identify this additional stellar mass dependence that was not apparent in the earlier work of \citetalias{Aird12}.

Our flexible, non-parametric approach also removes the need to assume an underlying functional form \citep[e.g. the power-law function assumed by \citetalias{Aird12};][]{Azadi15,Wang17} and thus enables us to recover the shape of the underlying distribution, informed directly by the data, as a function of stellar mass and redshift.
We find that \Psar\ generally has a double power-law shape at $\sar\gtrsim10^{-3}$, with a sharp turnover at the highest accretion rates and evolving toward higher \Sar\ at higher $z$. 
The earlier observational study of \citet{Bongiorno12} found indications of a high \Sar\ turnover in \Psar, while \citet{Aird13} suggested a double power-law shape (with a steep high-\Sar\ slope) was required to reconcile the measurements of \Psar\ and the galaxy stellar mass function with the observed shape of the AGN luminosity function \citep[see also][]{Caplar15,Bongiorno16}. 
Our study confirms these findings and provides new observational constraints on the overall shape and evolution of \Psar, independent of any assumed functional form. 
Our measurements also indicate a flattening or turnover of \Psar\ at the lowest accretion rates ($\sar\lesssim 10^{-2}-10^{-3}$), as discussed in Section~\ref{sec:discuss_sf} above.

By dividing our galaxy sample into star-forming and quiescent galaxies (based on their position in the SFR-\Mstel\ plane) and measuring \Psar\ within each population as a function of stellar mass and redshift, we are able to identify potential differences in the triggering and fuelling mechanisms of AGNs across the galaxy population. 
Earlier studies also found evidence that the normalization of \Psar\ is higher in galaxies with blue colours (e.g. \citetalias{Aird12}) or with higher SFRs \citep[e.g.][]{Azadi15}. \citet{Georgakakis14} also showed that the absolute space density as a function of \Sar\ is higher for AGNs with star-forming host galaxies than for those with quiescent host galaxies (separated based on the rest-frame UVJ colour criteria, e.g. \citealt{Williams09}), and found initial indications of differences in the shapes of the \Sar\ distributions. 
More recently, \citet{Wang17} measured \Psar\ directly in samples of red, green and blue galaxies (separated using dust-corrected rest-frame colours) in the GOODS-North and -South fields.
Consistent with our study, they found that the normalization of \Psar\ is higher in blue (i.e.~star-forming) galaxies than in red (i.e. quiescent) galaxies, has a different slope (fitting with a simple power-law function), and may evolve differently with redshift.
In particular, \citet{Wang17} identified a strong evolution of \Psar\ with redshift in red galaxies and a correspondingly high duty cycle in high-redshift ($z\sim2$) quiescent galaxies that is confirmed by our measurements.  
However, by adopting a larger sample of galaxies---covering a broader range in redshift and stellar mass---and exploiting new techniques to push the limits of the \textit{Chandra} survey data, we are able to identify a stellar mass dependence \citep[that was not found by][]{Wang17} and place tighter constraints on the shape and evolution of \Psar\ within both star-forming and quiescent galaxies.

\citet{Bongiorno16} and \citet{Georgakakis17} both use alternative approaches to probe the distribution of \Sar\ as a function of stellar mass and redshift.
We start with a sample of \emph{galaxies} and directly measure the AGN content of subsets of these galaxies using deep X-ray imaging \citep[see also \citetalias{Aird12};][]{Azadi15,Wang17}, whereas \citet{Bongiorno16} and \citet{Georgakakis17} start from a sample of \emph{X-ray selected AGNs} and measure the stellar masses of the host galaxies of this sample.
\citet{Bongiorno16} take the X-ray AGN sample from the XMM-COSMOS survey \citep{Hasinger07,Brusa10} and attempt to find parameterizations for the distribution of \Sar\ and the host galaxy stellar mass function that, after accounting for the X-ray selection effects, are consistent with the observed X-ray AGN sample.
Direct comparisons with our measurements are difficult as this method does not constrain the absolute normalization of \Psar.
Nonetheless, the basic patterns identified by \citet{Bongiorno16}---a double power-law shaped distribution of \Sar that shifts to higher \Sar\ at both higher redshifts and lower stellar masses---are consistent with our measurements (see e.g. Figure~\ref{fig:avl_vs_z}), although the parametrization adopted by \citet{Bongiorno16} lacks the low-\Sar\ turnover identified using our data-driven method.

\citet{Georgakakis17} adopt a larger sample of X-ray AGNs (combining data from a wide range of \textit{Chandra} and \textit{XMM-Newton} surveys) and use a non-parametric approach to find a \Psar\ that, when combined with independent measurements of the overall galaxy stellar mass function \citep[taken from][]{Ilbert13}, are consistent with the observed redshifts, luminosities, and host stellar masses of the X-ray selected sample. 
Despite the substantial differences between these studies,\footnote{As well as a distinct, independently developed methodology, \citealt{Georgakakis17} use data from a number of wide \textit{XMM-Newton} surveys (in addition to the \textit{Chandra} surveys used here), different multiwavelength imaging, and independent measurements of host galaxy stellar masses.}
the measurements of \Psar\ are in remarkably good agreement (see appendix~B of \citealt{Georgakakis17} for a direct comparison).
Both of our studies find a broad distribution of \Sar\ with a sharp cutoff at the highest accretion rates ($\sar\gtrsim0.1-1$) and indications of a flattening or turnover at very low accretion rates ($\sar\lesssim10^{-3}-10^{-2}$). 
As in \citet{Georgakakis17}, we identify a significant increase in the AGN duty with both increasing redshift and stellar mass.
While \citet{Georgakakis17} attribute the redshift evolution to an increase in the normalization of \Psar, our measurements reveal a shift of the distributions toward higher \Sar\ at higher $z$ (although some level of evolution in normalization cannot be ruled out). 
Our approach, extracting X-ray data from all galaxies in a sample to probe below the nominal sensitivity limits of \textit{Chandra}, allows us to place improved constraints at lower \Sar\ and across a wider range of galaxy stellar mass.
In addition, we measure \Psar\ independently as a function of stellar mass and redshift in star-forming and quiescent galaxies. 
Separately probing \Psar\ in star-forming and quiescent galaxies, as in our study, provides crucial insights into the different AGN fuelling mechanisms that appear to drive black hole growth in these different galaxy types.

\subsection{Average AGN accretion rates and the AGN main sequence}
\label{sec:averagecomp}

\refone{
A number of recent studies have used X-ray stacking techniques to measure \emph{average} AGN accretion rates for samples of galaxies as a function of redshift, stellar mass or SFR \citep[e.g.][]{Mullaney12b,Chen13,Delvecchio15,Yang17}.
As discussed by \citet{Hickox14}, stacking can account for the long-term variability in AGN accretion rates and reveal relationships between the average black hole growth rates and the relatively slowly varying properties of galaxies.
In contrast, our measurements of \Psar\ track the \emph{distribution} of AGN accretion rates, directly tracing the variability of AGN activity and revealing how these overall distributions depend on galaxy properties. 
X-ray stacking is equivalent to averaging over these distributions, losing much of the detailed information (such as the fraction of galaxies with black holes growing above a given accretion rate i.e.~the duty cycle).
Here, we briefly discuss how our findings relate to these prior studies.
}

\refone{
\citet[hereafter M12]{Mullaney12b} stacked the X-ray emission from samples of star-forming galaxies at $z\sim1$ and $z\sim2$ and identified a relation between the average X-ray luminosity and stellar mass with slope of $\sim$unity, designating this relation as a hidden ``AGN main sequence".}
\defcitealias{Mullaney12b}{M12}
\refone{As noted by \citetalias{Mullaney12b}, this relation is primarily due to higher stellar mass galaxies that are accreting at the same \emph{specific} accretion rates producing higher observed X-ray luminosities -- effectively the stellar mass selection bias identified by \citetalias{Aird12} and accounted for in this paper by measuring the distribution of \Sar\ directly.
}

\begin{figure}
\includegraphics[width=\columnwidth,trim=40 40 40 40]{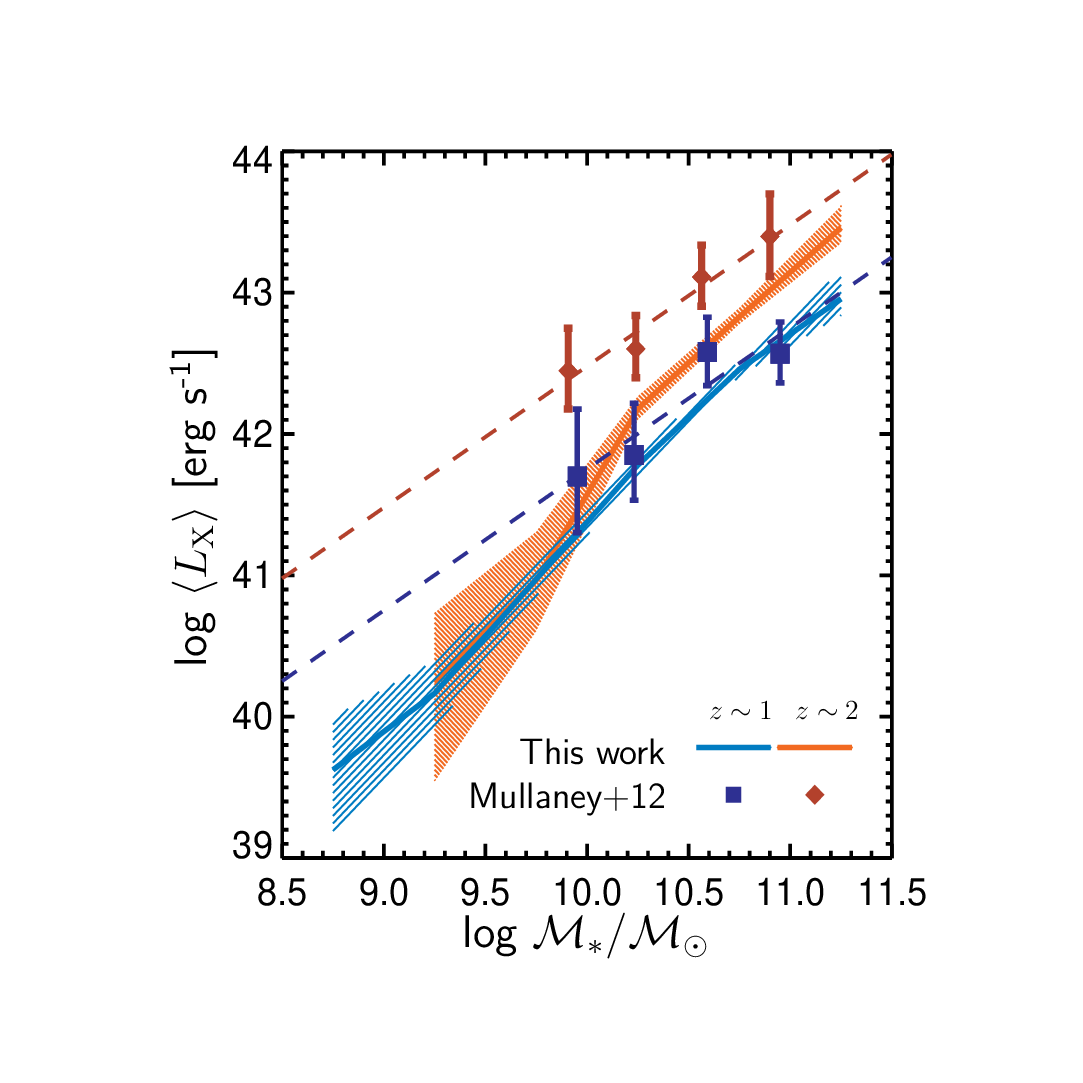}
\caption{
\refone{
Measurements of the ``AGN main sequence" from \citet{Mullaney12b} (red squares and blue diamonds indicating $z\sim1$ and $z\sim2$, as indicated) compared to estimates based on this work, averaging over our measurements \Psar\ for star-forming galaxies (coloured regions indicating 90 per cent confidence interval).
The dashed lines indicate linear fits to the \citet{Mullaney12b} measurements (with a fixed slope of 1), extrapolated over the entire stellar mass range.
Our estimates are consistent with a mild suppression relative to this linear relation at the lowest stellar masses (see Section~\ref{sec:averagecomp} for full discussion). 
}
}
\label{fig:agnmainsequence}
\end{figure}

\refone{
In Figure~\ref{fig:agnmainsequence} we estimate the mean X-ray luminosity of star-forming galaxy samples, $\langle \lx \rangle$, as a function of stellar mass and redshift using our measurements of \Psar, integrating over the full range of \Sar\ and converting back to $L_\mathrm{X}$ using Equation~\ref{eq:sar}. 
Our results (shaded regions) are consistent with a rising $\langle \lx \rangle$ with \Mstel\ i.e. an ``AGN main sequence". 
Our results agree with the \citetalias{Mullaney12b} measurements at $z\sim1$, although  our estimates lie below the extrapolation of a linear fit to the \citetalias{Mullaney12b} measurements (dashed blue line) at lower stellar masses \citep[see also][]{Yang17}.
This suppression of AGN activity at lower stellar masses is related to the reduction in the AGN duty cycle that we identify above (see Figure~\ref{fig:dutycycle_vs_z}), although the strength of the suppression of $\langle \lx \rangle$ is somewhat counteracted by the \emph{increase} in the average specific accretion rate, $\langle \sar\ \rangle$, of those AGNs (see Figure~\ref{fig:avl_vs_z}).}

\refone{
At $z\sim2$ our estimates of $\langle \lx \rangle$ in Figure~\ref{fig:agnmainsequence}
 lie systematically below the \citetalias{Mullaney12b} measurements of $\langle \lx \rangle$, most likely due to differences in the corrections for the contribution from star formation to the X-ray luminosity (primarily from high-mass X-ray binaries).
\citetalias{Mullaney12b} apply corrections based on locally-calibrated scaling relations \citep{Ranalli03}, whereas we apply a larger correction based on our work in \PaperI\ that indicates a larger X-ray luminosity per unit SFR at these redshifts \citep[see also][]{Lehmer16}.
Our measurements still reveal a clear ``AGN main sequence" at $z\sim2$, although there are indications that the slope may be steeper than unity due to a suppression of AGN activity at lower stellar masses. 
}

\refone{
\citet{Georgakakis17} also present estimates of $\langle \lx \rangle$ based on their independent measurements of \Psar. 
In contrast to our work, \citet{Georgakakis17} find a significantly \emph{flatter} slope in the relation between $\log \langle \lx \rangle$ and $\log \mstel$, associated with a suppression in AGN activity at \emph{high} stellar masses (cf. our suppression at \emph{low} stellar mass seen in Figure~\ref{fig:agnmainsequence} and steeper slope). 
However, \citet{Georgakakis17} do not separate the star-forming and quiescent galaxy populations and thus this flatter relation is likely associated with the lower accretion rates of AGN in quiescent galaxies (that start to dominate galaxy samples at $\log \mstel/\msun \gtrsim 10.5$ and $z\lesssim 1$), as revealed in our direct measurements of \Psar\ in the separated star-forming and quiescent galaxy populations (e.g. Figure~\ref{fig:pledd_sf_and_qu}).
}

\refone{
Other studies have measured the stacked X-ray emission from samples of galaxies as a function of SFR (rather than stellar mass), finding a positive correlation that could indicate a \emph{global} connection between the build of galaxies and the growth of their central black holes (e.g.~\citealt{Chen13}; see also \citealt{Hickox14}). 
However, whether this correlation is due to the underlying relationship between stellar mass and SFR in galaxies (i.e. the galaxy main sequence) and thus whether stellar mass or SFR are the primary driver of AGN activity remains an area of active study \citep[e.g.][]{Vito14,Delvecchio15,Azadi15,Yang17}.
Such work is further complicated by the presence of AGNs in quiescent galaxies (with very low SFRs) that may be fuelled by different physical mechanisms, as suggested in this paper \citep[see also][]{Kauffmann09}. 
Detailed measurements of \Psar\ as a function of both stellar mass and SFR are deferred to Paper~III. 
}

\subsection{AGN activity in dwarf galaxies}
\label{sec:lowmass}

\refone{At the lower redshifts probed by our study ($z<1.5$), our galaxy sample covers very low stellar masses corresponding to the dwarf galaxy regime ($\mstel/\msun \lesssim 3\times10^9$).
The existence, masses and growth rates of massive black holes
at the centres of such low-mass galaxies have been the focus of extensive study and interest 
\citep[e.g.][]{Greene07b,Reines13,Mezcua16,Pardo16,Baldassare16,Baldassare17,Hood17,Chen17}.
Such galaxies may provide the conditions under which the seeds of supermassive black holes can form and thus may provide local analogues that can help shed light on how supermassive black holes formed and subsequently grew in the very early ($z\gtrsim6$) universe \citep[e.g.][]{Volonteri10b,Reines16}.
}

\refone{
Our measurements place important constraints on both the black hole occupation fraction (the fraction of galaxies with a central, massive black hole) and the extent of AGN activity within dwarf galaxies.
Figure~\ref{fig:dutycycle_vs_z} indicates an AGN duty cycle of $f(\sar >0.01) \approx 0.1-0.9$~per~cent in galaxies with $8.5<\log \mstel/\msun<9.5$, which is in agreement with prior studies of X-ray AGNs in dwarf galaxy samples (e.g.~\citealt{Pardo16} find an AGN fraction of $\sim$0.6~--~3~per~cent in galaxies with $\mstel\le 3\times 10^9\msun$ to a luminosity threshold of $\lx>10^{41}$~\ergs, comparable to our \Sar\ limit).
Our measurements are also consistent with no evolution with redshift in the AGN duty cycle for dwarf galaxies out to at least $z\sim1$, in contrast to higher mass galaxies where there is significant evolution.
Despite this low overall AGN duty cycle, we \emph{do} identify a population of high accretion rate ($\sar\gtrsim0.1$) AGNs within our dwarf galaxy samples, as indicated by our measurements of \Psar\ (see e.g. Figure~\ref{fig:pledd_all}). 
Thus, a small fraction of dwarf galaxies ($\sim0.1$ per cent) must contain massive black holes that are growing at high accretion rates relative to their stellar mass. 
Indeed, our measurements of \Psar\ at high \Sar\ are consistent across the full range of galaxy masses probed in this study, with significant differences only becoming apparent at lower \Sar\ (see e.g. the two lowest redshift panels of Figure~\ref{fig:pledd_zbins} for all galaxies or star-forming galaxies).
}

\refone{
It is also important to note, especially for this low-mass regime, that our measurements place constraints on the distribution of \emph{specific} accretion rates: the levels of black hole growth relative to the \emph{total galaxy stellar mass}. 
The scaling between the black hole mass and the host galaxy mass may have significantly larger scatter or break down completely in these very low-mass galaxies \citep[e.g.][]{Greene08,Greene10,Kormendy13,Graham15,Reines15}.
Thus, our measurements of \Psar\ may not reflect the underlying distribution of \emph{Eddington ratios}; in fact, our measurements track the convolution of the probability distribution of Eddington ratios and the distribution of black hole masses at a given galaxy stellar mass.
Our measurements are also degenerate with the overall black hole \emph{occupation} fraction: 
the low AGN duty cycle that we find could indicate that a large fraction of galaxies in this mass regime simply lack central massive black holes and thus would never be observed as an AGN \citep[see also][]{Trump15}.
It is not possible with the current data to break such degeneracies, although our measurements do provide a lower limit on the occupation fraction and show that actively accreting massive black holes \emph{can} exist in such low-mass galaxies.
Thus, despite these limitations, our measurements of \Psar\ provide important observational constraints on the incidence of AGNs and the extent of black hole growth within the dwarf galaxy population.
}

\subsection{AGN accretion rate distributions in cosmological simulations}
\label{sec:cosmosims}

\refone{
Recent large-volume cosmological hydrodynamic simulations \citep[e.g.][]{Dubois14,Vogelsberger14,Schaye15} are able to model the flow of gas into the very central regions of galaxies and---via sub-grid modelling---predict the accretion rates onto the central black holes.
Here we briefly discuss how the results of recent cosmological simulations compare with our measurements of \Psar\ and AGN duty cycles. 
}

\refone{
\citet{Volonteri16} presented predictions for both Eddington ratio distribution functions and distributions of \Sar\ (accretion rate scaled relative to the total predicted galaxy stellar mass) based on the Horizon-AGN simulation. 
The predictions are in good agreement with our observational results for galaxies with moderate stellar masses, $\mstel\gtrsim10^{10}\msun$, revealing a broad, roughly power-law distribution at moderate accretion rates ($\sar\sim10^{-3}-0.1$), a break at higher accretion rates ($\sar \gtrsim 0.1-1$), and a shift toward higher \Sar\ at higher redshifts. 
The simulation also predicts the low \Sar\ flattening or turnover that is seen in our measurements. 
At lower stellar masses, the simulation tends to over-predict \Psar, possibly indicating the need for stronger supernova-driven feedback in the simulations that would limit the fuelling AGN activity in lower mass galaxies (see discussion in \citealt{Volonteri16} and \citealt{Dubois15}).
}

\refone{
\citet{Rosas-Guevara16} presented predictions for the observed properties of AGNs in the EAGLE simulation \citep[see also][]{McAlpine17}. 
They find that the average Eddington ratio for the simulated galaxy sample increases with redshift, consistent with our observed trends for $\langle \sar \rangle$ (see Figure~\ref{fig:avl_vs_z}). 
The average Eddington ratio is found to decrease at higher black hole masses at a fixed redshift, which 
appears broadly consistent with our measurements of $\langle \sar \rangle$ seen in Figure~\ref{fig:avl_vs_z} (left panel) but may be in conflict with the \emph{increase} in the overall duty cycle at higher \Mstel\ that we observe (see Figure~\ref{fig:dutycycle_vs_z}). 
Further, detailed comparisons of EAGLE predictions of \Psar\ that can be directly compared to our measurements are required to reconcile these issues and assess the  modelling of AGN accretion in EAGLE.}

\refone{
\citet{Sijacki15} presented predicted Eddington ratio distributions as a function of redshift based on the Illustris simulation. 
Consistent with our observations of \Psar, the predicted distributions exhibit a broad range of accretion rates at any epoch.
The distributions shift toward higher accretion rates at higher redshifts and narrow as they reach the Eddington limit, as seen in our measurements.
\citet{Sijacki15} also predict that average Eddington ratios are roughly constant over a broad range of black hole mass (at a given redshift) but drop very rapidly for the very highest mass black holes ($\mathcal{M}_\mathrm{BH} \gtrsim 10^{9}-10^{10} \msun$, depending on redshift). 
This mass dependence differs from the much more gradual decrease in average Eddington ratio with increasing black hole mass found by \citet{Rosas-Guevara16} in the EAGLE simulation and the changes in the duty cycle and $\langle \sar \rangle$ with stellar mass that we measure. Direct comparisons of the Illustis predictions of \Psar\ as a function of \Mstel\ are required to explore this issue.}

\refone{In conclusion, our measurements of \Psar\ are generally consistent with
the broad range of accretion rates predicted by recent cosmological simulations and the overall evolution of these distributions, shifting to higher accretion rates at higher redshifts. 
However, none of the simulations appear to produce the exact stellar mass dependence of \Psar\ that we measure.
Directly comparing our measurements with predications of \Sar\ (i.e.~relative to host stellar mass) as a function of galaxy type, stellar mass and redshift will enable crucial tests of new and ongoing cosmological simulations, providing a more refined observational comparison than generic characterizations of the AGN population (e.g.~luminosity functions).
}

\section{Conclusions}
\label{sec:conclusions}

\renewcommand{\Psar}{$p(\log \sar \giv \mstel,z)$}
In this paper, we use deep \textit{Chandra} X-ray data to probe the distribution of AGN accretion rates within large samples of stellar-mass-selected star-forming and quiescent galaxies out to $z\sim4$.
Our main conclusions are as follows:

\begin{enumerate}[leftmargin=*,itemsep=5pt]\item
Our measurements show that \Psar\ has a broad distribution that is roughly consistent with a power-law with a steep cutoff at the highest accretion rates ($\sar\gtrsim0.1-1$) and a flattening or possibly a turnover at low accretion rates ($\sar\lesssim10^{-3}-10^{-2}$).
These broad distributions are consistent with stochastic processes being primarily responsible for the fuelling of AGNs and substantial variability in the level of black hole growth over short timescales relative to the host galaxy.

\item
This broad distribution of accretion rates is found for both the star-forming and quiescent galaxy populations, across a wide range in stellar mass and out to $z\sim4$. 
At higher stellar masses ($\mstel\gtrsim10^{10}\msun$) and lower redshifts ($z\lesssim2$) we are able to identify significant differences in \Psar\ in star-forming versus quiescent galaxies (at fixed \Mstel\ and $z$), whereby the distribution in quiescent galaxies appears shifted to lower \Sar. Thus, quiescent galaxies tend to host weaker (lower \Sar) AGNs and the AGN duty cycle (the fraction of galaxies with AGN above a fixed limit of $\sar>0.01$) is generally lower.

\item
We find that the \Psar\ evolves strongly with redshift for star-forming galaxies with $\mstel\gtrsim10^{10}\msun$, shifting toward higher \Sar\ at higher redshifts and leading to a significant increase in the AGN duty cycle.
This evolution may be due to the increased availability of cold gas at higher redshifts that increases both the frequency and luminosities of accretion events at earlier cosmic times. 
However, the distributions appear to be truncated at a point roughly corresponding to the Eddington limit, indicating that black holes may ultimately self-regulate their own growth at high redshifts when copious gas is available. 

\item 
\refone{Our results indicate that \Psar\ in star-forming galaxies depends on stellar mass.
The probability of a lower mass star-forming galaxy hosting a low-accretion-rate AGN ($\sar \lesssim 0.1$) is suppressed compared to within higher mass galaxies. 
The evolution of the AGN duty cycle with redshift is thus weaker at lower stellar masses.
We measure a low duty cycle ($\lesssim 0.9$ per~cent) in the lowest mass, dwarf galaxy regime ($\mstel\lesssim 3\times 10^{9} \msun$) and find no evidence for evolution out to $z\sim1$.
Nonetheless, high accretion rate AGNs ($\sar \gtrsim 0.1$) are identified in a small fraction ($\sim0.1$~per~cent) of dwarf galaxies.
}

\item
In quiescent galaxies, we also find strong evolution whereby \Psar\ shifts to higher \Sar\ at higher $z$ for $\mstel\gtrsim10^{10}\msun$. 
The duty cycle appears to \emph{decrease} at the highest stellar masses, $\mstel \gtrsim 10^{11}\msun$ (in contrast to the \emph{increase} with stellar mass found for star-forming galaxies).
We also find that the duty cycle for quiescent galaxies evolves strongly with redshift and is comparable to star-forming galaxies at $z\sim2$.
The rapid evolution of the duty cycle may also be related to the increased availability of cold gas at higher redshifts or could be due to younger (recently quenched) stellar populations with increased stellar mass loss.

\end{enumerate}

Our new measurements have enabled a detailed mapping of the distribution of black hole growth over the galaxy population.
While our results support a picture of stochastically driven AGN activity, our measurements show that such activity is enhanced at higher redshifts, in high-mass star-forming galaxies, and in high-redshift quiescent galaxies. 
\updd{
These findings indicate that black hole growth (via AGN activity) does not simply track galaxy growth (via star formation) and that the triggering and fuelling of AGNs may be driven by different physical mechanisms across the galaxy population (e.g. in star-forming versus quiescent galaxies). 
Our detailed measurements  place strong constraints on the extent of these different AGN fuelling mechanisms over a wide range of stellar mass and redshift.
}

\vspace{15pt}
\refone{The measurements of accretion rate probability distributions, \Psar, and the AGN duty cycle, $f(\sar>0.01)$, presented in this paper are made available in machine-readable format at \url{http://doi.org/10.5281/zenodo.1009605}.}

\section*{acknowledgements}
We thank the anonymous referee for helpful comments that contributed to this paper.
We acknowledge helpful conversations with Dave Alexander, Francesca Civano, Andy Fabian, Ryan Hickox, James Mullaney, Kirpal Nandra and Marta Volonteri.
JA acknowledges support from ERC Advanced Grant FEEDBACK 340442. 
AG acknowledges the {\sc thales} project 383549 that is jointly funded by the European Union  and the  Greek Government  in  the framework  of the  programme ``Education and lifelong learning''. 
This work is based in part on observations taken by the CANDELS Multi-Cycle Treasury Program and the 3D-HST Treasury Program (GO 12177 and 12328) with the NASA/ESA HST, which is operated by the Association of Universities for Research in Astronomy, Inc., under NASA contract NAS5-26555.
The scientific results reported in this article are based to a significant degree on observations made by the \textit{Chandra} X-ray Observatory.

\appendix
\section{Two-component (galaxy+AGN) fitting of spectral energy distributions and AGN-corrected estimates of star formation rates}
\label{app:agnfast}

\begin{figure}
\includegraphics[width=\columnwidth,trim=10 10 10 0]{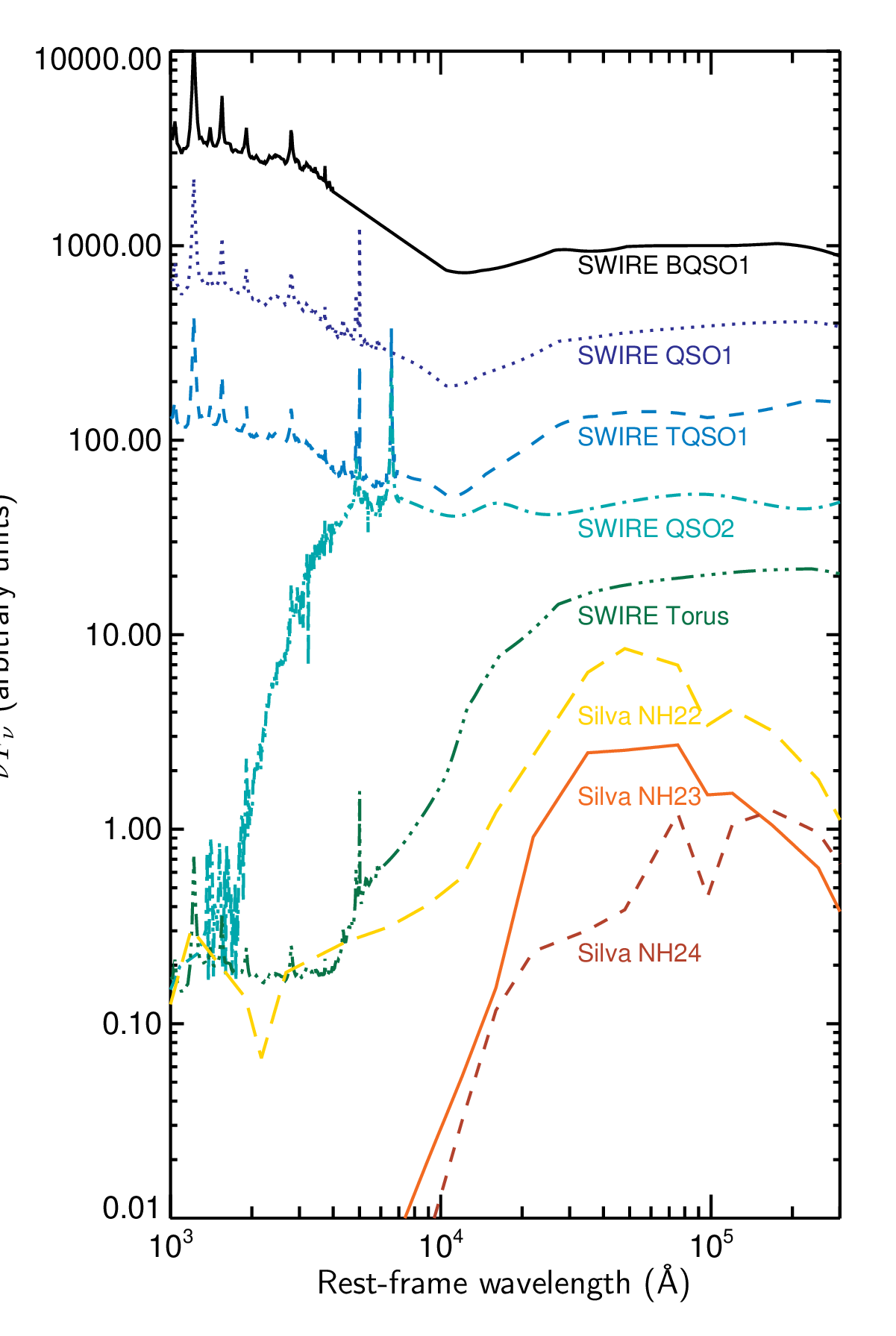}
\caption{The empirical AGN templates (arbitrarily re-scaled) that are adopted in our two-component SED fitting. The top five templates are taken from the SWIRE template library \citep{Polletta07}, while the bottom three templates are from \citet{Silva04} and allow for an AGN contribution that dominates in the IR but produces negligible optical or UV emission. }
\label{fig:agn_templates}
\end{figure}

  \begin{figure*}
\includegraphics[width=\columnwidth,trim=50 45 20 20]{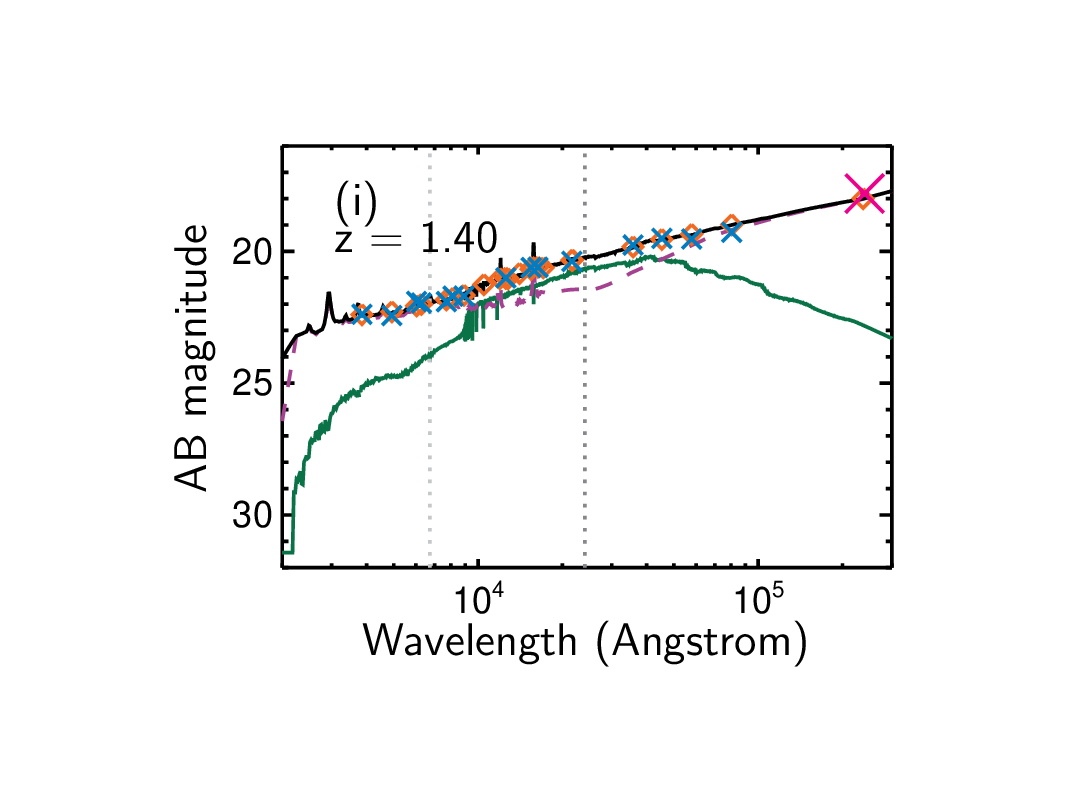}
\includegraphics[width=\columnwidth,trim=50 45 20 20]{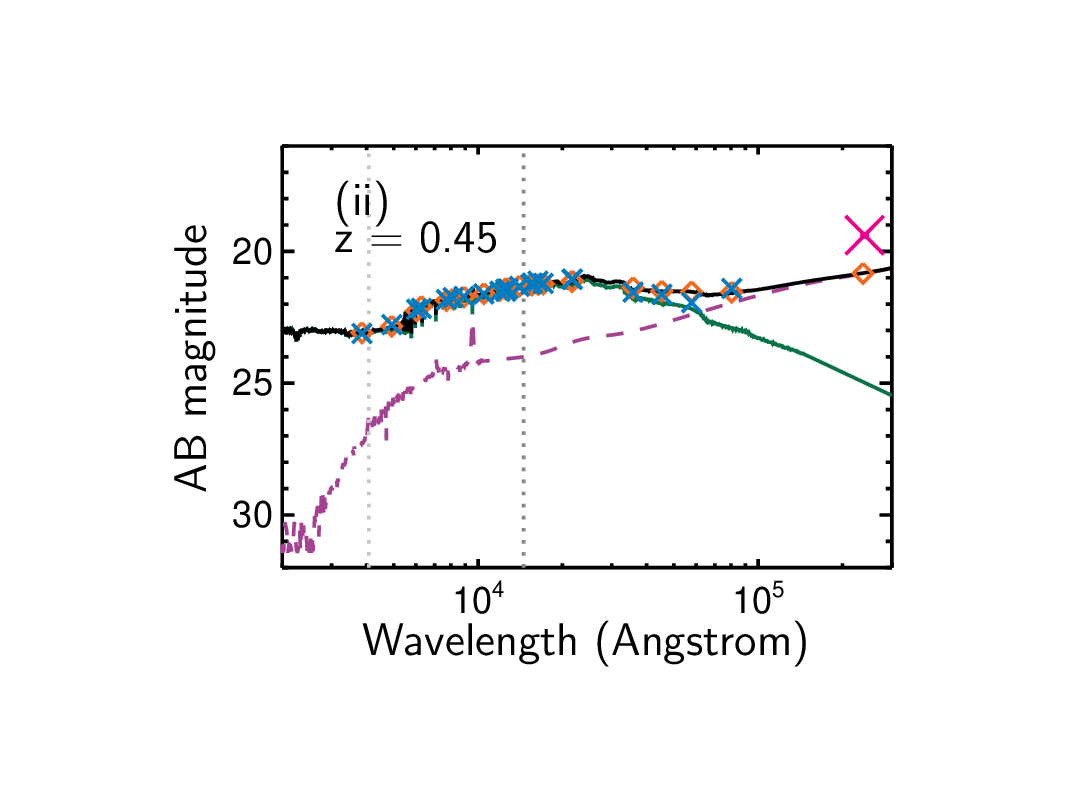}
\includegraphics[width=\columnwidth,trim=50 30 20 20]{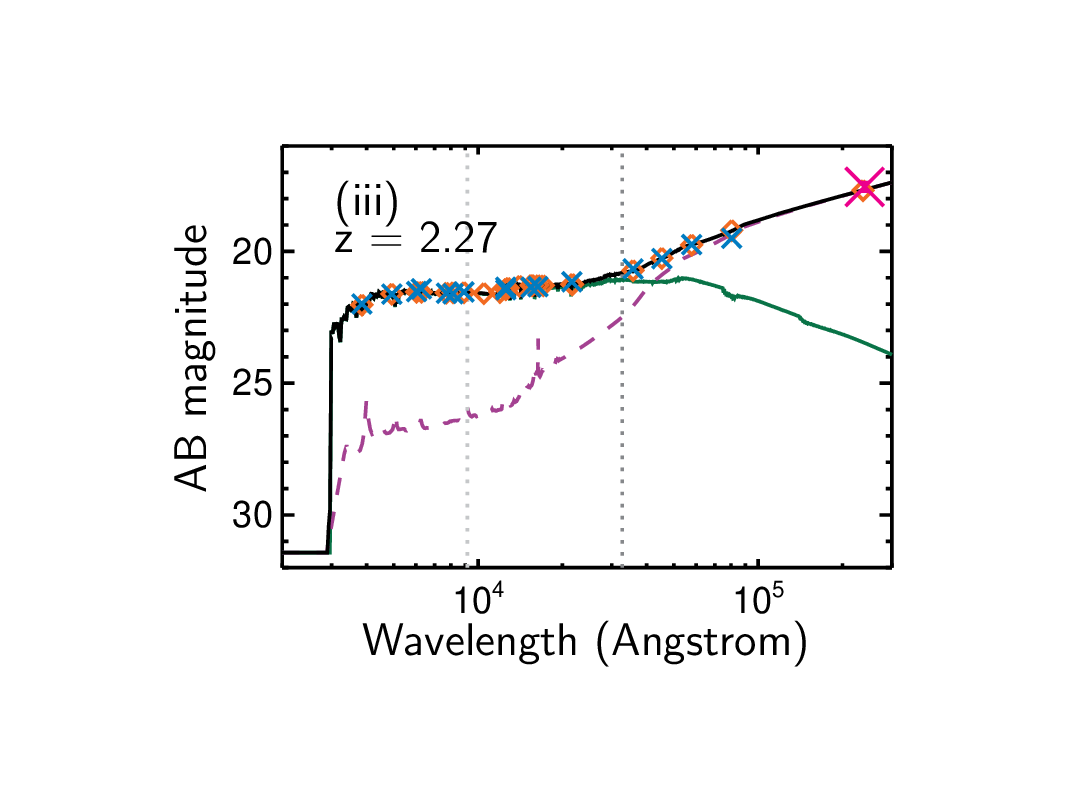}
\includegraphics[width=\columnwidth,trim=50 30 20 20]{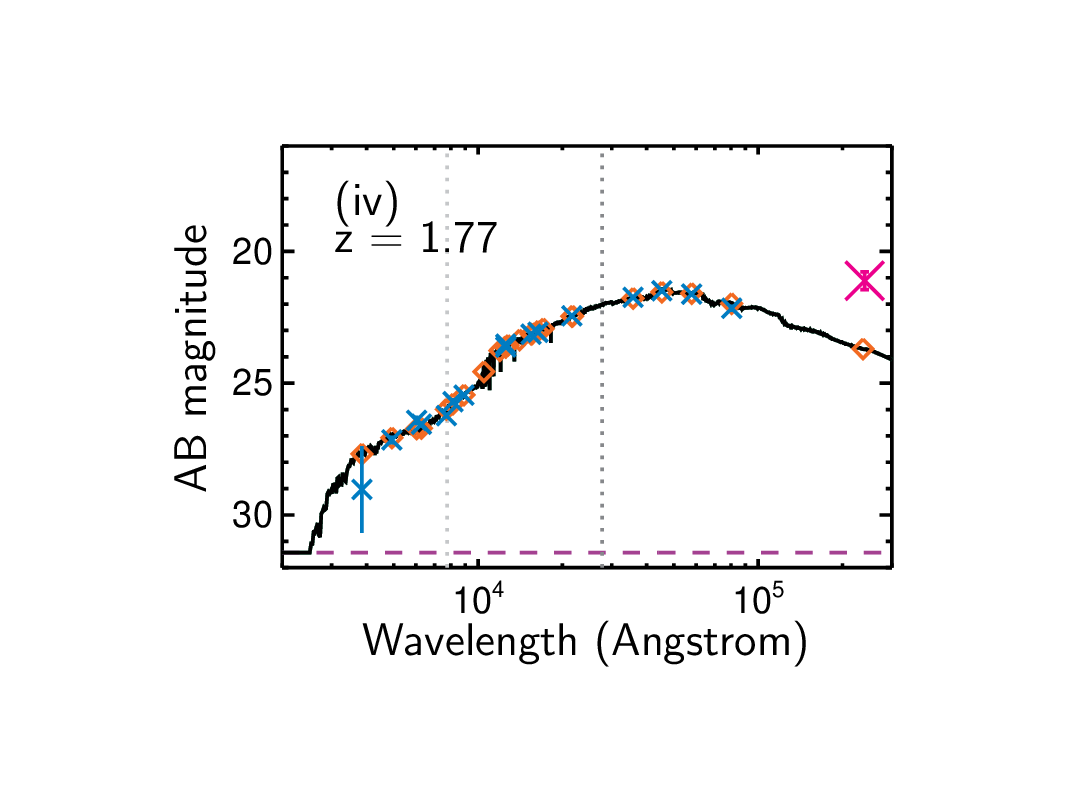}
\caption{Example SEDs and results of our two-component (AGN+galaxy) fitting.
Blue crosses indicate the observed photometry (and errors) and the orange diamonds indicate the predicted photometry with the best-fitting two-component model. The green solid and purple dashed lines indicate the best-fitting galaxy and AGN templates respectively; the solid black line is the sum of the two components. 
The pink crosses indicate the observed MIPS 24\micron\ flux, which is \emph{not} considered in the SED-fitting but may be used to estimate the SFR.
Panel (i) shows an example of an AGN-dominated source. The host galaxy makes a contribution around $\sim2\times10^{4}$\AA\ (rest-frame $\sim 1$\micron) that allows us to estimate the stellar mass. All of the 24\micron\ emission can be associated with the AGN component. Our estimates of the SFR for such sources are highly uncertain and thus we exclude these AGN-dominated sources when dividing our sample based on their SFRs (i.e. into star-forming and quiescent galaxies).  
Panel (ii) shows a source with evidence of an AGN contribution in the MIR ($\sim$3--8\micron), but the extrapolation of the AGN template does not account for all of the observed 24\micron\ flux; the SFR is estimated from the sum of the UV based on the best-fitting galaxy template (green line) and the IR emission traced by the 24\micron\ flux after subtracting the AGN contribution. 
Panel (iii) is a source with a clear power-law AGN contribution in the MIR, although the galaxy light dominates at shorter wavelengths. All of the 24\micron\ emission can be associated with the AGN component, thus we estimate the SFR based on the galaxy component of the SED fit. 
Panel (iv) shows a source where the UV-to-MIR SED is best described by a galaxy template (without any AGN contribution). The excess at 24\micron\ is associated with the star formation (the dust emission is not included in the galaxy model templates). We adopt the UV+IR estimate of the SFR, without any correction for an AGN contribution.
}
\label{fig:sedexamples}
\end{figure*}

In appendix~A of Paper~I we describe our approach to estimate the SFRs of galaxies, based on either summing the UV+IR emission or from fitting the UV-to-MIR spectral energy distributions (SEDs) with stellar population synthesis (SPS) models using a modified version of the FAST code \citep{Kriek09}. 
The SED fitting is also used to estimate the total galaxy stellar mass, \Mstel. 
However, for sources with an AGN, the host galaxy light can be severely contaminated by the AGN emission, especially at UV and MIR wavelengths. 
In this appendix, we describe how we correct for potential AGN contamination by including AGN templates in our SED fitting and SFR estimates, implementing a two-component (galaxy+AGN) fitting approach \citep[see also e.g.][]{Bongiorno07,Bongiorno12,Lusso12,Rovilos14,Ciesla15}.

We allow for a \emph{potential} AGN contribution for any galaxy with a significant X-ray detection \citep[i.e. detected with a false probability $<4\times10^{-6}$ in the full, soft, hard or ultra-hard \textit{Chandra} energy bands, see Section~\ref{sec:data} above and][]{Laird09}.
Allowing for a potential AGN contribution for every galaxy in our overall sample, regardless of X-ray properties, is computationally prohibitive and risks systematically altering our SFR and \Mstel\ estimates when we know that the majority of galaxies are not expected to have a significant AGN contribution.
There will be AGN that are missed by our X-ray selection \citep[in particular, luminous but obscured sources that may be revealed by their MIR colours but are not seen at X-ray wavelengths e.g.][]{Donley12,Mendez13}. 
However, such sources are expected to constitute a small fraction of our overall galaxy sample and thus,
while the SFR and \Mstel\ esitimates may be inaccurate for individual sources, they will not have a significant impact on our overall results. 

We adopt a library of eight, empirically determined templates for the SEDs of AGNs (that exclude any host galaxy contribution) by combining the five AGN-dominated templates from the \citet{Polletta07} SWIRE template library (namely, the Torus, TQSO1, BQSO1, QSO1 and QSO2 templates) with the three composite SEDs of X-ray selected AGNs with absorption column densities $\mathrm{N_H}=10^{22-23}, 10^{23-24}$ and $10^{24-25}$~cm$^{-2}$ from \citet{Silva04}.
These templates are chosen to provide a reasonable sampling of the range of possible AGN SEDs.
The templates from \citet{Silva04} are adopted as they correspond to optically obscured AGN with very red, power-law emission that can dominate at IR wavelengths (with negligible UV-to-optical emission). 
We do not attempt to estimate $\mathrm{N_H}$ for our X-ray selected AGN sample and thus do not apply any prior preference for a particular template based on the X-ray absorption or any other data when fitting the SED of a given source. 
Our AGN template library is shown in Figure~\ref{fig:agn_templates}.

We allow for linear combinations of each of our eight AGN templates with each of our grid of galaxy templates (described in Paper~I), such that
\begin{equation}
f_\mathrm{total}(\lambda) = A_1 f_\mathrm{gal}(\lambda) + A_2 f_\mathrm{AGN}(\lambda)
\end{equation}
where $f_\mathrm{gal}(\lambda)$ and $f_\mathrm{AGN}(\lambda)$ represent a given galaxy and AGN SED template, respectively, and $A_1$ and $A_2$ are unknown, free parameters corresponding to the relative scalings of the two templates. 
For each possible combination (i.e. for every galaxy template combined with each of the eight AGN templates), we calculate $A_1$ and $A_2$ using $\chi^2$ minimization to fit to the observed photometry. 
We also determine the $\chi^2$ for every galaxy template without any AGN component (i.e. fixing $A_2=0$). 
We retain the standard FAST template error function when calculating $\chi^2$, although we truncate the uncertainty to 50~per cent (in flux) at long wavelengths (where the standard template error function becomes large, mostly to allow for uncertainties related to an AGN contribution that we now model directly). 
We idenitfy the minimum $\chi^2$ over the entire grid of galaxy and AGN combinations, which we retain as our best-fitting estimate of the overall SED.
Following our approach in Paper~I, we also calculate \emph{a posteriori} estimates of galaxy properties (i.e. $\mathrm{SFR_{SED}}$, \Mstel) by marginalizing over the full grid of possible galaxy and AGN combinations (as well as the ``no-AGN" possibility), although we only take the single best $\chi^2$ (and corresponding $A_1$ and $A_2$) for a given combination rather than performing a fully Bayesian marginalization.
We use the same approach to provide a best \emph{a posteriori} estimate of the fractional AGN contribution, $F_{AGN,5000}$, defined as the fraction of the light at 5000\AA\ that is attributed to the AGN template component.
We flag sources where $F_{AGN,5000}$ is greater then 50 per cent as AGN-dominated in the UV-optical.

The next step of our analysis is to determine our best estimate of the SFR for each galaxy in the X-ray detected sample. 
In Paper~I, we describe our ``SFR ladder'' for galaxies: when a galaxy is detected at 24\micron\ we adopt $\mathrm{SFR_{UV+IR}}$;
for galaxies without 24\micron\ detections we adopt $\mathrm{SFR_{SED}}$. 
Here, we adopt a similar approach but apply a correction for any AGN contribution. 
First, we extrapolate the best-fitting (minimum $\chi^2$) AGN template to the infrared and estimate the expected flux in the MIPS 24\micron\ band due to the AGN component. 
We subtract this expected AGN contribution from any observed flux at 24\micron.
If the source is still detected at $>3\sigma$ after subtracting the AGN contribution, we calculate $\mathrm{SFR_{UV+IR}}$ using equation A1 of Paper~I but subtracting the expected AGN contribution to both the 24\micron\ flux and $L_{\nu 2800}$ based on our two-component SED fitting. 
For sources where all of the 24\micron\ flux is associated with the AGN component, or sources without 24\micron\ detections, we adopt $\mathrm{SFR_{SED}}$ as the best estimate of the SFR. 
Figure~\ref{fig:sedexamples} shows several examples of our two-component SED fits for a range of sources with differing levels of AGN contribution.
In panels i and iii all of the observed 24\micron\ flux can be accounted for with the AGN template and thus we estimate the SFR based on the best-fitting SPS model for the galaxy component (the SFR estimate will be highly uncertain in the case of panel i, where the observed SED is dominated by the AGN component). In panels ii and iv the extrapolation of the AGN component is not suffiicient to account for the 24\micron\ flux (the galaxy models do not include the far-IR emission from star-formation-heated dust that is needed to account for this emission). 
In such cases we estimate the SFR from the UV+IR emission (via equation A1 of Paper~I), subtracting any AGN contribution.

\begin{figure}
\includegraphics[width=\columnwidth,trim=35 50 10 25]{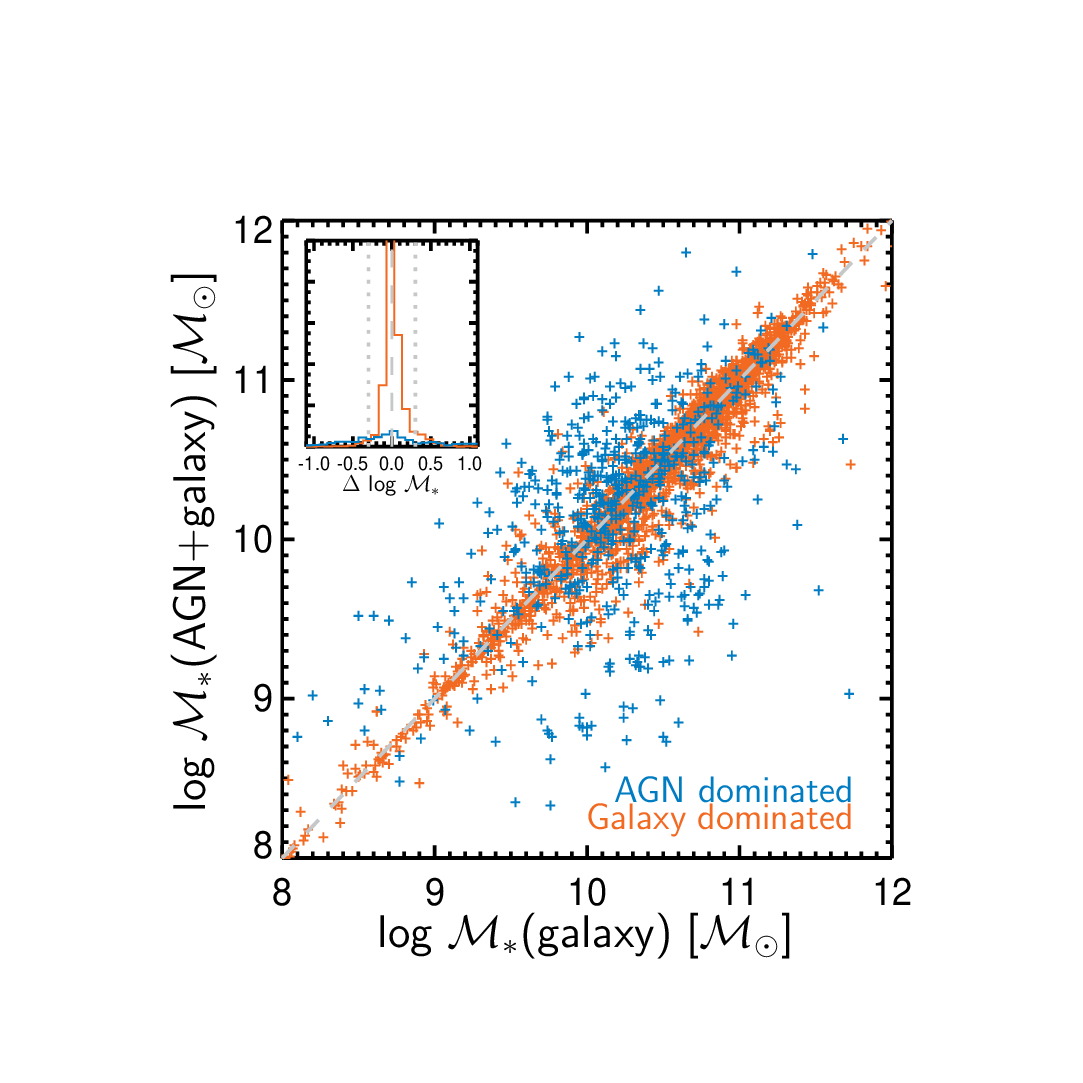}
\includegraphics[width=\columnwidth,trim=35 10 10 0]{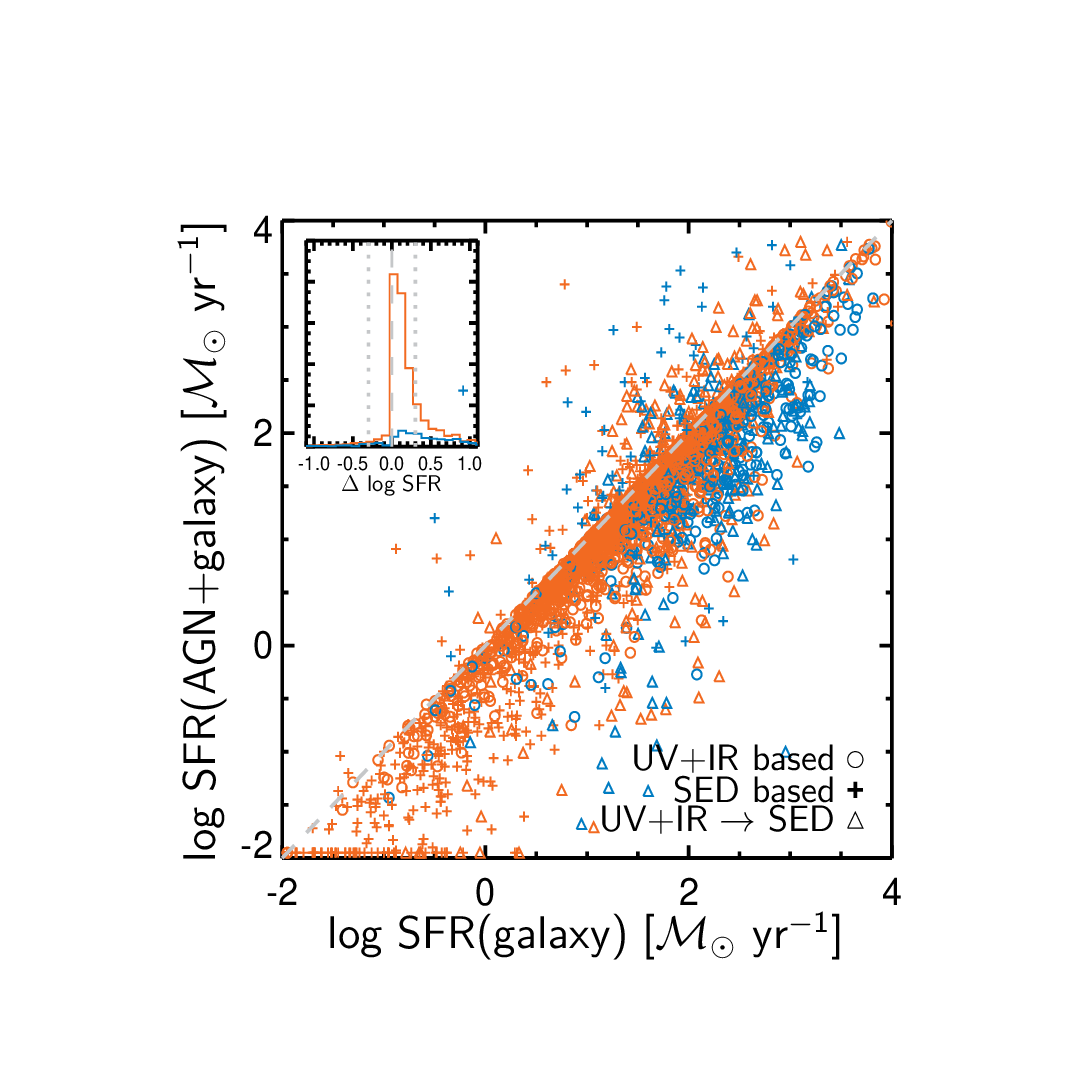}
\caption{Comparison of the stellar mass (top panel) and SFR (bottom panel) estimates when considering the galaxy contribution to the SED only (x-axes) and performing a two-component fitting with both AGN and galaxy contributions (y-axes). 
Points in blue correspond to AGN-dominated sources (more than 50 per cent of the light at 5000\AA\ is attributed to the AGN in the two-component SED fits), whereas orange points indicate that the optical light is dominated by the galaxy component. 
In the bottom panel circles indicate sources where UV+IR SFRs are used for both the SFR(galaxy) and SFR(AGN+galaxy) estimates; plus symbols indicate sources that are not detected at 24\micron\ (and thus the SED-based estimates of the SFR are used in both cases); triangles indicate sources that are detected at 24\micron\ but all of the flux can be attributed to the AGN component, thus we move from the UV+IR estimates to the SED-based estimates of the SFR. Inset plots show histograms of the logarithmic diffence between the two estimates. The dashed grey line is at 0 whereas the dotted grey lines indicate $\pm0.3$~dex.
}
\label{fig:agn_vs_noagn}
\end{figure}

In Figure~\ref{fig:agn_vs_noagn} we compare our estimates of \Mstel\ and SFR based on our two-component fitting to the previous (galaxy only, single-component) approach.
The stellar masses (top panel) generally change by $\lesssim0.3$~dex for galaxy-dominated sources ($F_\mathrm{AGN,5000}<50$ per cent, shown in orange) and there is no overall systematic offset. For AGN-dominated sources (blue points, $\sim$ 20 per cent of the X-ray selected sample) there is a larger scatter, although our two-component fitting should provide an improved estimate of the stellar mass for such sources.
For the SFRs (bottom panel) there is a clearer systematic shift for both AGN- and galaxy-dominated sources that reduces our estimates by $\sim0.1-0.3$~dex on average and by up to an order of magnitude in extreme cases, demonstrating the importance of applying corrections for the AGN contribution to the UV, optical and IR light.

\refone{
An updated version of the FAST code, implementing the changes described here and in appendix~A of Paper~I, is made available at \url{https://github.com/jamesaird/FAST}.
}

\section{Bayesian mixture modelling of the distribution of AGN accretion rates and corrections for the contribution of the host  galaxy}
\label{app:updatedbayes}

In this paper we adopt the flexible Bayesian mixture modelling approach---described in detail in appendix~B of Paper~I---to estimate $p(\log \sar \giv \mstel,z)$, the intrinsic distribution of \Sar\ for galaxies of a given stellar mass, redshift, and galaxy classification (e.g. star-forming or quiescent). 
Here, we describe two further refinements of the methodology for the present study: 
1) adapting the method to recover the distribution of specific black hole accretion rates, \Sar, rather than X-ray luminosities; and 
2) accounting for (and thus removing) the expected contribution to the observed X-ray flux from galactic (non-AGN) processes.

The first amendment---converting from X-ray luminosity to \Sar ---simply requires us to replace all \LX\ terms in appendix~B of Paper~I with \Sar\ and adjust the scaling factor for each galaxy, $k_i$, that converts this quantity to a predicted number of X-ray counts.
Thus, following Equation~\ref{eq:sar} above that defines \Sar, 
equation B3 of Paper~I is adapted to
\begin{equation}
k_i  = \kappa \times \eta(z_i) \times t _i \times \dfrac{\mathcal{M}_i}{\msun}
\end{equation}
where $\kappa = 1.04 \times 10^{34}$ is a constant factor defined by our assumed (constant) bolometric correction and our scaling of \Sar\ to ``Eddington-ratio equivalent" units.
The scaling for an individual galaxy, $i$, also depends on the conversion factor, $\eta(z_i)$,\footnote{The $\eta(z_i)$ factor converts between rest-frame 2--10~keV luminosity and the observed count rate in the 2--7~keV \textit{Chandra} band. We assume an unabsorbed spectrum with photon index $\Gamma=1.9$ for every source. Thus, the conversion factor depends only on the redshift of the source, $z_i$.} the effective X-ray exposure at the galaxy position, $t_i$, and the total stellar mass of the galaxy,~$\mathcal{M}_i$.
We model $p(\log \sar \giv \mstel,z)$ as a sum of Gamma distributions, as described by equations B7-B10 of Paper~I (substituting \LX\ with \Sar).
Similar to Paper~I, we fix the positions of the Gamma distributions in \Sar, adopting a logarithmically spaced grid over the range $-5 \le \log \lambda_j \le 2$ in steps of 0.2~dex, where $\lambda_j$ is the scale parameter of each mixture component.
We also fix the shape parameter to $\alpha_j=3.0$ for every component and apply a prior that prefers a smooth variation between adjacent components. 
To ensure sufficient flexibility, we allow for an additional component at $\log \lambda_j=-7$ that accounts for galaxies that do not contribute to the observed distribution of \Sar\ and is not subjected to our smoothness prior.
The integral over all mixture components is required to sum to 1.
Our overall model for $p(\log \sar \giv \mstel,z)$  for a sample of galaxies in a given range of \Mstel\ and $z$ thus has 36 free parameters (linked via the smoothness prior), corresponding to the normalizations, $A_j$, of each mixture component.

Our second refinement is to apply a correction for the contribution from galactic (non-AGN) X-ray processes. 
In Paper~I, we found that the observed distributions of \LX\ for galaxies exhibit a narrow peak at $\lx \lesssim 10^{42}$~\ergs, associated with the emission from (predominantly) high- and low-mass X-ray binaries. We used this peak to trace the star-forming ``X-ray main sequence" and re-calibrate the scaling between X-ray luminosity and galaxy properties (SFR, \Mstel). 
In the present paper, we aim to measure the distribution of AGN activity, as traced by the X-ray emission, and thus need to remove this non-AGN contribution. 
We adjust the likelihood function from \PaperI\ for the data from a single galaxy to allow for three possible origins of the observed X-ray counts, each described by an independent Poisson process: 1) emission from the AGN, 2) emission from other processes in the galaxy, and 3) a background component. Thus, equation B4 of \PaperI\ can be re-written as
\begin{align}
\mathcal{L}(N_i \giv \sar, L_\mathrm{G}, b_i, t_i, z_i) = \nonumber\\
 \sum_{C_\mathrm{A}=0}^{N_i} \sum_{C_\mathrm{G}}^{N_i-C_\mathrm{A}} 
\Bigg(
			&	\frac{(k_i \sar)^{C_\mathrm{A}}}{C_\mathrm{A}!}e^{-k_i \sar} \nonumber\\
			& \frac{(l_i L_\mathrm{G})^{C_\mathrm{G}}}{C_\mathrm{G}!}e^{-\l_i L_\mathrm{G}} \nonumber\\
			&		\frac{b_i^{C_\mathrm{B}}}{C_\mathrm{B}!}e^{-b_i} \Bigg)
			\label{eq:newlik}
\end{align}
where $C_\mathrm{A}$, $C_\mathrm{G}$ and $C_\mathrm{B}$ are unknown, nuisance parameters that represent the integer X-ray counts due to the AGN, galactic processes, and the background, respectively. 
The summations in Equation~\ref{eq:newlik} are equivalent to marginalizing over the possible values of the nuisance parameters: $C_\mathrm{A}$, $C_\mathrm{G}$ and $C_\mathrm{B}$. 
We require
\begin{equation}
C_\mathrm{A} + C_\mathrm{G} + C_\mathrm{B} = N_i
\end{equation}
so that these three unknown parameters sum to the total observed counts, $N_i$, from galaxy $i$. 
The $k_i\sar$, $l_i L_\mathrm{G}$, and $b_i$ terms correspond to the (non-integer) underlying count \emph{rates} related to the AGN, galactic, and background components, respectively, which produce the \emph{integer} counts from each component ($C_\mathrm{A}$, $C_\mathrm{G}$ and $C_\mathrm{B}$) via a Poisson process. 
The background count rate, $b_i$, is well-determined and is taken from our background maps \citep[see Paper~I;][]{Georgakakis08}.
The likelihood function of Equation~\ref{eq:newlik} describes our knowlege of the AGN count rate ($k_i\sar$) for galaxy $i$, which is ultimately combined with the data from other galaxies in a sample to place posterior constraints on $p(\log \sar \giv \mstel,z)$.

The expected galaxy count rate, $l_i L_\mathrm{G}$, depends on a conversion factor,\footnote{The $l_i$ factor converts from a rest-frame 2--10~keV luminosity to observed 2--7~keV counts, assuming an unabsorbed, $\Gamma=1.9$ X-ray spectrum appropriate for the galactic emission. Thus $l_i=\eta(z_i)\times t_i$.} $l_i$, and the X-ray luminosity due to galactic processes, $L_\mathrm{G}$. 
We can apply a prior for $L_\mathrm{G}$ for a given galaxy based on information at other wavelengths: the SFR and \Mstel\ measured from the UV-to-IR SED (see Appendix~\ref{app:agnfast} above). 
To scale from SFR and \Mstel\ to an expected X-ray luminosity, we adopt Model~5 from Paper~I, which accounts for contributions from both high- and low-mass X-ray binaries that should roughly scale with SFR and \Mstel, respectively. 
To allow for uncertainty in the measured SFR and \Mstel, as well as the large uncertainty in the conversion to X-ray luminosity, we describe our prior using a single Gamma distribution for each galaxy,
\begin{equation}
\pi(L_\mathrm{G} \giv \tilde{\mathcal{M}_i}, \tilde{\mathrm{SFR}_i})
= \frac{1}{\Gamma(\alpha)\theta^{\alpha}} L_\mathrm{G}^{\alpha-1} 
e^{-L_\mathrm{G}/\theta}
\end{equation}
with a shape parameter ${\alpha=2}$ and a scale parameter ${\theta= \tilde{L_\mathrm{G}}/(\alpha-1)}$.
A Gamma distribution with $\alpha=2$ is roughly equivalent to a logarthmic ``1$\sigma$" uncertainty of $\sim0.8$~dex, thus allowing for a large uncertainty in the galactic X-ray luminosity, $L_\mathrm{G}$, for an \emph{individual} galaxy.  
We choose to model this uncertaintiy with a Gamma distribution (rather than a normal or lognormal distribution) as the Gamma distribution is conjugate to our Poisson likelihood function and thus simplifies our calculations. 
The scale parameter, $\theta$, is chosen such that the mode of the prior distribution corresponds to $\tilde{L_\mathrm{G}}$, our best estimate of the X-ray luminosity from galactic processes based on our measurement of the stellar mass ($\tilde{\mathcal{M}_i}$) and SFR ($\tilde{\mathrm{SFR}_i}$) for galaxy~$i$.

Our overall likelihood function for the data from all galaxies in a given stellar mass--redshift bin ($\mathbf{D}_\mathrm{bin}$) is then given by
\begin{align}
\mathcal{L}(\mathbf{D}_\mathrm{bin}) =\;\;\;\;\;\;\;\;\;\;\;\;&\nonumber\\
\prod_{i=1}^{n_\mathrm{source}} 
\int_0^\infty \int_0^\infty
	\bigg[&\mathcal{L}(N_i \giv \lambda, L_\mathrm{G}, b_i, t_i, z_i)	
               \; \pi(L_\mathrm{G} \giv  \tilde{\mathcal{M}_i}, \tilde{\mathrm{SFR}_i}) \nonumber\\
  	            & \pi(\lambda \giv \mathcal{M}_\mathrm{bin}, z_\mathrm{bin}) \bigg] \;
                dL_\mathrm{G} \;d\lambda
                \label{eq:likall}
\end{align}	
where $\lambda \equiv \sar$ and 
\begin{equation}
\pi(\lambda \giv \mathcal{M}_\mathrm{bin},z_\mathrm{bin}) \;d\lambda \equiv p(\log \sar \giv \mathcal{M}_*,z) \;d\log\sar
\end{equation}
which is the distribution of sBHAR that we model as a mixture of Gamma distributions, as discussed above. 
As in Paper~I, the overall likelihood function described by Equation~\ref{eq:likall} can be reduced to 
\begin{equation}
\mathcal{L}(\mathbf{D}_\mathrm{bin}) = 
\prod_{i=1}^{n_\mathrm{source}} \left[\sum_{j=1}^{K} A_j \mathcal{W}_{ij} \right]
\end{equation}
where $A_j$ are the normalizations of each mixture component in our model of $p(\log \sar \giv \mathcal{M}_*,z)$ and $\mathcal{W}_{ij}$ are pre-computed weights for each component $j$ and source $i$. The sum is taken over all $K$ mixture components and the product is taken over all $n_\mathrm{source}$ galaxies with a given \Mstel, $z$, and classification.

This revised scheme effectively allows for an additional ``background" X-ray emission for every galaxy in our sample based on the expected galactic emission (allowing for an uncertainty in the luminosity of such emission). 
By modelling the galactic emission separately from the AGN X-ray luminosity we are able to remove this component and recover the underlying distribution of \Sar\ related to the AGN population within our galaxy samples (removing the low-luminosity peaks found in Paper~I).
In practice, for galaxies where we directly detect a luminous X-ray AGN ($\lx \gtrsim 10^{42}$~\ergs) the expected galactic contribution is usually negligible.
For the large number of galaxies in our samples without direct X-ray detections, but where we still use the available X-ray information, this correction is more important.
Furthermore, for high-mass, high-redshift, or high-SFR galaxies the galactic X-ray emission can be comparable to or exceed a weak AGN. 
Nevertheless, the results of this paper are mostly based on the broad, high-luminosity tail of the luminosity distribution (related to AGN). We generally do not consider ranges of \Sar\ where the galactic X-ray emission is expected to be the dominant component and our measurements of 
$p(\log \sar \giv \mathcal{M}_*,z)$, even after applying the correction described in this appendix, are uncertain.

\section{The impact of AGN-dominated sources on measurements of the accretion rate probability distribution}
\label{app:agndom}

\begin{figure*}
\includegraphics[width=0.94\textwidth,trim=0 25 0 30]{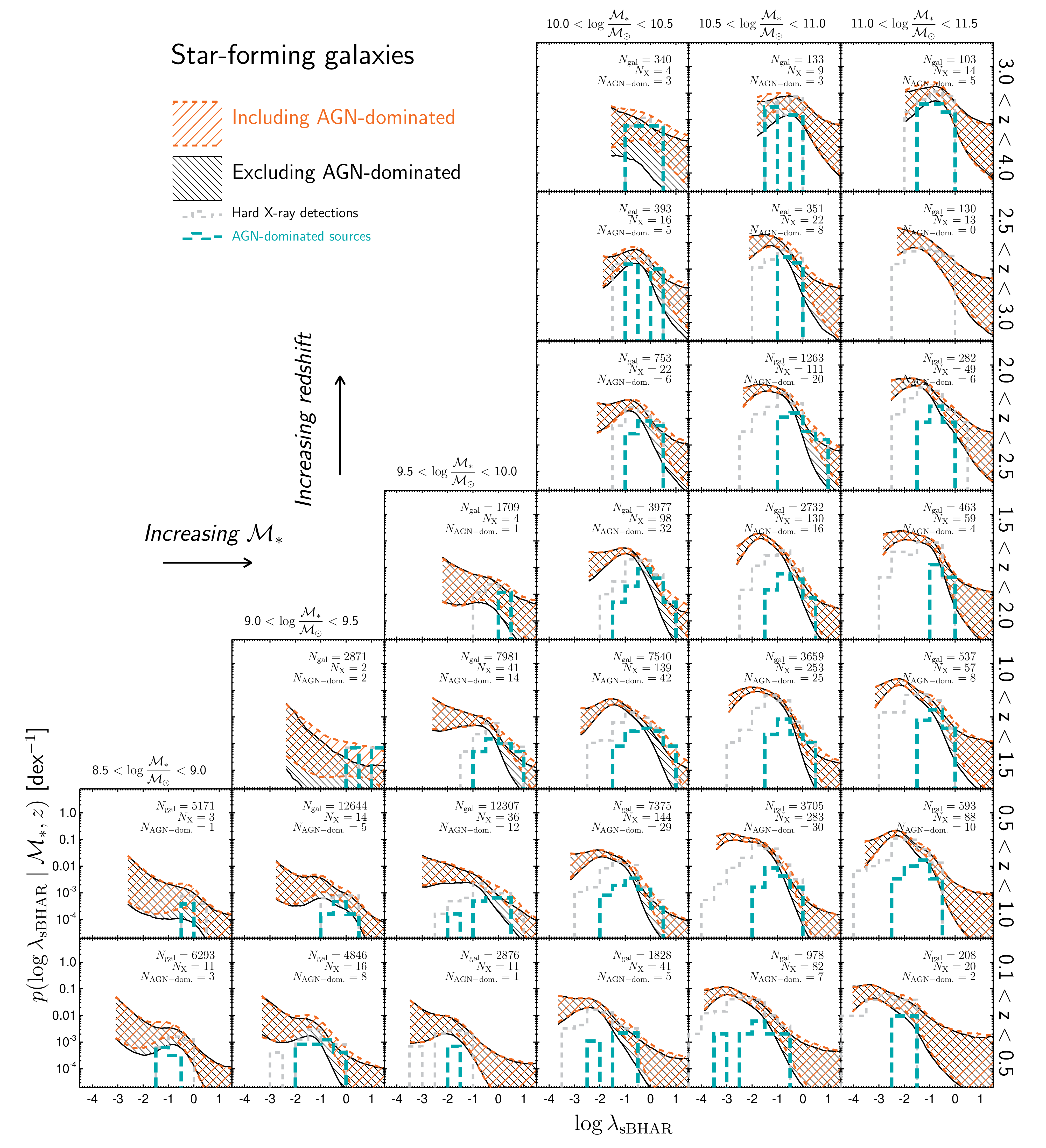}
\caption{
Comparison of measurements of \Psar\ for star-forming galaxies when including \emph{all} AGN-dominated sources in the star-forming galaxy sample (orange hatched regions) and when excluding these sources (black hatched regions, corresponding to the results in the main paper, see Figure~\ref{fig:pledd_sf_and_qu}), where the regions indicate the 90 per cent confidence intervals on our measurements.
Grey histograms indicate the observed distribution of \Sar\ for sources with X-ray detections (excluding AGN-dominated sources), without any correction for X-ray incompleteness, whereas the green histograms show the observed distribution of \Sar\ for AGN-dominated sources. 
While AGN-dominated are associated with the highest \Sar, excluding them does not have a significant impact on our measurements. 
}
\label{fig:pledd_sf_incexc}
\end{figure*}

\begin{figure*}
\includegraphics[width=0.94\textwidth,trim=0 25 0 30]{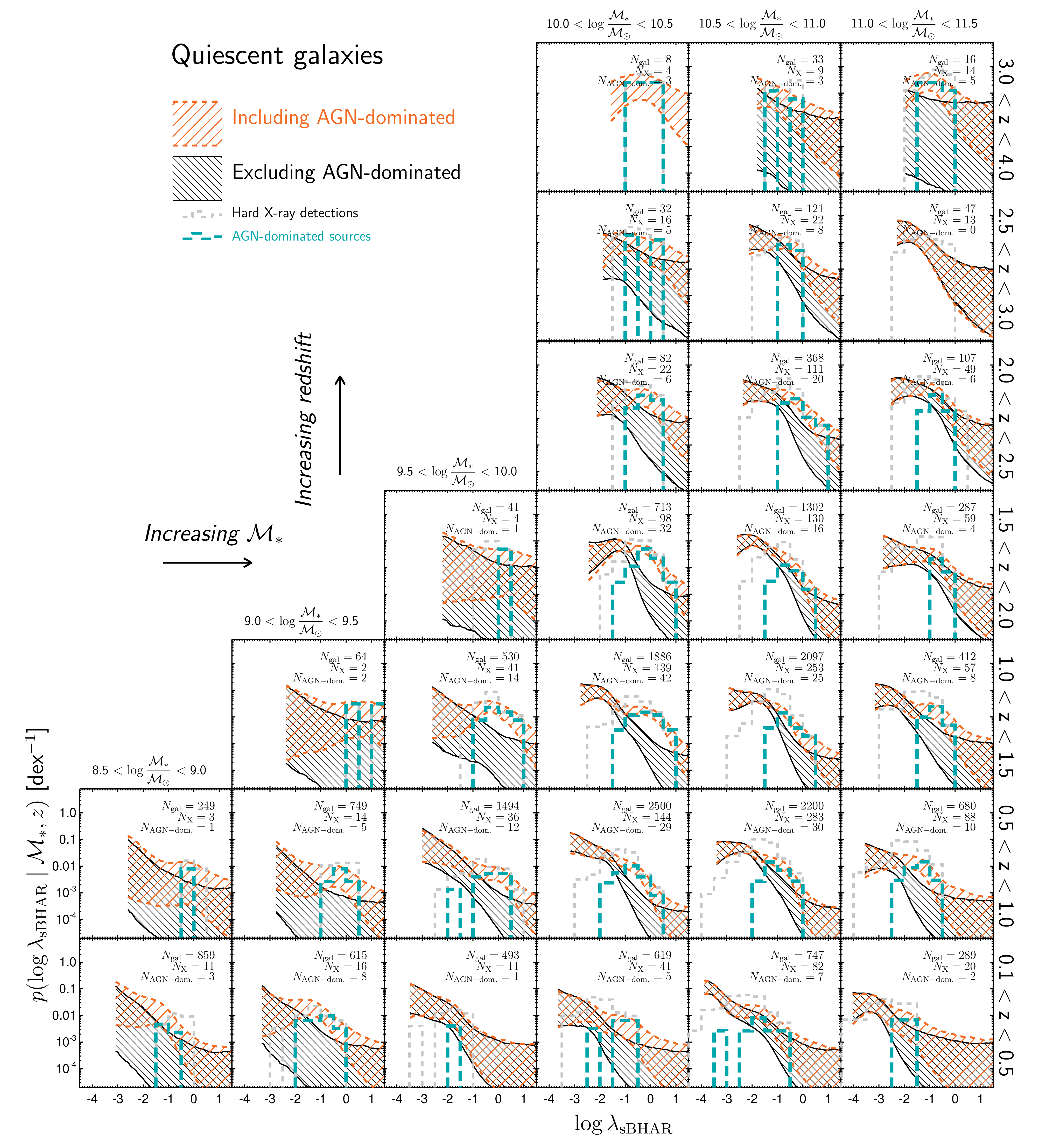}
\caption{
Comparison of measurements of \Psar\ for quiescent galaxies when including \emph{all} AGN-dominated sources in the quiescent galaxy sample (orange hatched regions) and when excluding these sources (black hatched regions, corresponding to the results in the main paper, see Figure~\ref{fig:pledd_sf_and_qu}).
Given the small sample sizes and the lack of high \Sar\ sources, assigning \emph{all} AGN-dominated sources to the quiescent galaxy samples has a significant impact on our results.
However, we note that all AGN-dominated sources residing in quiescent galaxies is an extreme and unlikely scenario that is reflected by the complex shapes of the orange regions in this figure (see text of Appendix~\ref{app:agndom} for discussion).
}
\label{fig:pledd_qu_incexc}
\end{figure*}

Via our SED-fitting (see Appendix~\ref{app:agnfast}), we are able to identify roughly 20 per cent of our X-ray detected sample ($\lesssim 0.3$ per cent of the overall galaxy sample) as ``AGN-dominated", where the light associated with the AGN component dominates over any galaxy component at optical wavelengths (specifically, $>$50 per cent of the light at rest-frame 5000~\AA\ is associated with the AGN component in the SED fit). 
For such sources, it is extremely difficult to accurately measure the SFR of the galaxy (the UV-to-MIR SED template is generally poorly constrained and in many cases the 24\micron\ light is also dominated by the AGN component), precluding a classification of the host galaxy as star-forming or quiescent (see Figure~\ref{fig:sfr_vs_mstel}). 
However, as the galaxy component tends to constitute a significant fraction of the light at rest-frame $\sim$1\micron\ ($>$ 50 per cent of the light at 1\micron\ is associated with the galaxy component in 77 per cent of the AGN-dominated sources), we can still obtain reasonably accurate estimates of the host stellar mass \citep[see also][]{Georgakakis17}. 
We thus choose to \emph{include} AGN-dominated sources in our measurements of \Psar\ for all galaxies, but \emph{exclude} these sources when dividing our galaxy into star-forming and quiescent (see Sections~\ref{sec:all}--\ref{sec:sfqu}). 
In this appendix, we examine the impact of excluding or including the AGN-dominated sources on our measurements.

In Figure~\ref{fig:pledd_sf_incexc} we compare estimates of \Psar\ for star-forming galaxies as a function of stellar mass and redshift when excluding AGN-dominated sources (black hatched regions, corresponding to our main results presented in Figure~\ref{fig:pledd_sf_and_qu}) and when \emph{all} sources identified as AGN-dominated are included in the star-forming galaxy sample (orange hatched regions).
The grey histograms correspond to the observed distribution of \Sar\ for X-ray detected sources in star-forming galaxies (neglecting any correction for X-ray completeness and excluding AGN-dominated sources), whereas the green histograms indicate the observed distributions for the AGN-dominated sample. 
We note that including \emph{all} of the AGN-dominated sources in the star-forming galaxy sample is conservative as some of these sources may be hosted by quiescent galaxies.
However, for the majority of the stellar masses and redshifts probed in this study the star-forming galaxies dominate the total number density.
Our results also indicate an increase in the probability of hosting an AGN for star-forming galaxies (compared to quiescent galaxies) and independent studies have also found that AGN (both Type-1 and Type-2 sources) are primarily hosted by star-forming galaxies \citep[e.g.][]{Rosario12,Rosario13,Stanley15}. 
Thus, it is reasonable to assume that the majority of the AGN-dominated sources in our sample are in fact hosted by star-forming galaxies.

It is immediately apparent in Figure~\ref{fig:pledd_sf_incexc} that AGN-dominated sources are preferentially associated with the highest \Sar\ sources at all stellar masses and redshifts. 
Assuming that AGN-dominated sources correspond to Type-1, unobscured AGNs, this trend could indicate an association between \Sar\ (and thus, Eddington ratio) and the levels of obscuration \citep[see also e.g.][]{Raimundo10,Merloni14}. 
However, to be identified as AGN-dominated with our method requires a high AGN luminosity relative to the host and thus a high \Sar. 
An unobscured, ``Type-1" source with a low \Sar\ would not be classified as AGN-dominated.  
The association between higher \Sar\ and identification as an AGN-dominated source is  a natural consequence of our method.
A full investigation of the relationship between obscuration (at optical or X-ray wavelengths) and \Sar\ is beyond the scope of this work.

The prevalence of AGN-dominated sources at higher \Sar\ introduces a small, systematic shift in our estimates of \Psar, with the orange hatched regions in Figure~\ref{fig:pledd_sf_incexc} being systematically higher at $\sar \gtrsim0.1$ than when the AGN-dominated sources are excluded (black hatched regions). 
However, in general the 90 per cent confidence intervals overlap, indicating that the difference is not significant. 
Even at high \Sar, the AGN-dominated sources tend to constitute at most $\sim$50 per cent of our X-ray AGN sample. 
Our measurements track the broad shape of the probability distribution over logarithmic scales; 
thus, excluding the AGN-dominated population does not have a significant effect on these measurements. 
Significant discrepancies (i.e. no overlap between the 90 per cent confidence regions) are seen at the highest accretion rates for the $10<\log \mstel/\msun<10.5$ stellar mass bin and both the $1.0<z<1.5$ and $1.5<z<2.0$ redshift bins, although the shape of the distribution remains the same and these differences do not alter any of the conclusions we draw in this paper. 
We thus conclude that our measurements of \Psar\ for the star-forming galaxy sample (as a function of \Mstel\ and $z$) are not significantly affected by the exclusion of the AGN-dominated sources and retain the measurements that exclude these sources as our best estimates presented in Figure~\ref{fig:pledd_sf_and_qu} and used for subsequent analysis. 

We note that a similar result (i.e. no significant impact from excluding AGN-dominated sources) is found when considering all galaxies.
However, as we are still able to constrain the stellar mass of these sources we choose to adopt the measurements that \emph{include} the AGN-dominated sources in Figure~\ref{fig:pledd_all} and subsequent analysis of the ``All galaxies" sample.

In Figure~\ref{fig:pledd_qu_incexc} we perform a similar analysis but considering the quiescent galaxy samples. 
We assign \emph{all} of the AGN-dominated sources (that cannot be classified individually) to the quiescent galaxy sample for the ``Including AGN-dominated" estimates (orange hatched regions). 
As discussed above, we would expect the vast majority of AGN-dominated sources to actually lie in star-forming host galaxies, thus this approach is extremely conservative and provides strict upper limits on \Psar\ for the quiescent galaxy population. 
Figure~\ref{fig:pledd_qu_incexc} shows that including all AGN-dominated sources in the quiescent galaxy sample can have a significant impact on our measurements, mostly due to the relatively small size of our quiescent galaxy samples and the lack of the high \Sar\ sources within the original sample.
Nonetheless, in some bins---given the large uncertainties on the original ``Excluding AGN-dominated" measurements---the differences are not significant.  
In a number of bins (e.g. $1.0<z<1.5$ and $10.0<\log\mstel/\msun<10.5$) including AGN-dominated sources substantially warps the recovered shape of the probability distribution, possibly indicating that the AGN-dominated sources are drawn from a distinct population. 
The overall impact of including the AGN-dominated sources is to bring the probability distributions into closer agreement with the measurements for star-forming galaxies.

We thus conclude that excluding AGN-dominated sources from our measurements of \Psar\ for star-forming galaxies does not have a significant impact on our results or conclusions. 
The impact of AGN-dominated sources (primarily, but not exclusively, expected to be Type-1 AGN) on our measurements of \Psar\ for quiescent galaxies is less clear, but we believe our measurements are robust given the reasonable assumption that the majority of the AGN-dominated sources are in fact hosted by star-forming galaxies.

\label{lastpage}
\end{document}